\documentclass{article}
\usepackage[utf8]{inputenc}
\usepackage{amsmath}
\usepackage{geometry}
\usepackage{float}
\usepackage{mathrsfs}
\usepackage{amssymb}
\usepackage{dsfont}
\usepackage{xcolor}
\usepackage{ntheorem}
\usepackage{graphicx}
\usepackage{caption}
\usepackage{listings}
\usepackage{enumerate}
\usepackage{natbib}
\usepackage{hyperref}
\usepackage{booktabs}
\title{
\textbf{Identifying  average causal effect in regression discontinuity design with auxiliary data}}
\author{Xinqin Feng*\\
    \small Department of Probability and Statistics, Peking University\\
    \small fengxinqin@pku.edu.cn\\
    Wenjie Hu*\\
    \small Perelman School of Medicine, University of Pennsylvania\\
    \small wenjie.hu@pennmedicine.upenn.edu\\
    Pu Yang\\
    \small Department of Information and Computing Science, Peking University\\
    \small yang\_pu@pku.edu.cn\\
    Tingyu Li\\
    \small Chongqing Key Laboratory of Child Neurodevelopmental and Cognitive Disorders,\\ \small Children's Hospital of Chongqing Medical University, \\\small Ministry of Education Key Laboratory of Child Development and Disorders, \\\small National Clinical Research Center for Child Health and Disorders \\
    \small tyli@vip.sina.com \\
    and \\
    Xiao-Hua Zhou\\
   \small Beijing International Center for Mathematical Research and Department of Biostatistics,\\\small
   Peking University\\
   \small azhou@math.pku.edu.cn}
\newcommand{\I}{\mathbb{I}}
\newcommand{\ind}{\perp \!\!\! \perp}
\newcommand{\abs}[1]{\left|#1 \right|}
\newcommand{\norm}[1]{\left\|#1\right\|}

\newtheorem{assumption}{Assumption}
\newtheorem{theorem}{Theorem}
\newtheorem{remark}{Remark}
\newtheorem{proposition}{Proposition}
\newtheorem{lemma}{Lemma}

\newtheorem{definition}{Definition}
\begin{document}
\date{}

\maketitle
\newpage
\begin{abstract}
    Regression discontinuity designs are widely used when treatment assignment is determined by whether a running variable exceeds a predefined threshold. 
    However, most research focuses on estimating local causal effects at the threshold, leaving the challenge of identifying treatment effects away from the cutoff largely unaddressed. 
    The primary difficulty in this context is that the treatment assignment is deterministically defined by the running variable, violating the commonly assumed positivity assumption.
    In this paper, we introduce a novel framework for identifying the  average causal effect in regression discontinuity designs.
    Our approach assumes the existence of an auxiliary variable for which the running variable can be seen as a surrogate, and an additional dataset that consists of the running variable and the auxiliary variable alongside the traditional regression discontinuity design setup. 
    Under this framework, we propose three estimation methods for the ATE, which resembles the outcome regression, inverse propensity weighted and doubly robust estimators in classical causal inference literature. Asymptotically valid inference procedures are also provided.
    To demonstrate the practical application of our method, 
    simulations are conducted to show the good performance of our methods; besides, we use the proposed methods to  assess the causal effects of vitamin A supplementation on the severity of autism spectrum disorders in children, where a positive effect is found but with no statistical significance.
\end{abstract}
\begin{center}\textit{Keywords:} Average treatment effect; Causal inference; Data fusion; Extrapolation; Regression discontinuity design.\end{center}
\newpage
\section{Introduction}\label{introduction}
The regression discontinuity design, initially proposed by \cite{thistlethwaite1960regression}, is currently one of the most widely used approaches for evaluating causal inference drawn from quasi-experimental data. 
In this design, each unit is allocated to treatment/control group based on the value of its running variable. 
Specifically, we focus on the sharp regression discontinuity design in this paper, wherein treatment assignment is known to depend deterministically on the running variable. 
For example, each unit receives treatment if its running variable exceeds a predetermined threshold. 

In this paper, we study the extrapolation of causal inference within the sharp regression discontinuity design, i.e., evaluating causal effects on broader populations. 
Our motivating example is the vitamin A supplementary program conducted by Children’s Hospital of Chongqing Medical University \citep{Xi20}. 
This program aimed to evaluate the efficacy of taking vitamin A on reducing the severity of autism spectrum disorders in Chinese children. 
In this program, each participant was assigned to receive vitamin A supplementation if their baseline serum retinol concentration fell below the normal level. 
This was exactly a regression discontinuity design in which the running variable was the serum retinol concentration.
In addition, the research team also collected another dataset containing the serum retinol concentration along with retinoic acid level \citep{Feng2024association}, which was understood in biological science that retinol must be converted to to participate in human physiological activities \citep{Shearer2012}.

Despite many methodological works on the regression discontinuity design, most of them have focused on causal effects in the subpopulation with the running variable at the threshold \citep{Hahn01,Ludwig07,Imbens12,armstrong2018optimal,calonico2018effect,Calonico19,imbens2019optimized,armstrong2020simple,Arai21}.
To evaluate causal effects in a broader population, previous work has essentially sought to incorporate additional information into the regression discontinuity design. 
There are primarily three approaches to incorporate additional information. 
The first approach is to directly impose assumptions on the conditional expectations of potential outcomes given the running variables. 
For example, this may involve restricting conditional expectations within a specific function space \citep{dong2015identifying,sun2023extrapolating,cattaneo2017comparing,lee2022regression}. 
However, these assumptions are often overly restrictive, particularly in sharp regression discontinuity designs, since neither medical knowledge nor the data itself can support them. 
The second approach is to modify the canonical design to tackle the non-positivity issue. 
This includes methods such as comparative regression discontinuity designs \citep{wing2013strengthening,wing2018regression} and multi-cutoff designs \citep{Cattaneo21-2,Bertanha20,zhang2022safe}. 
However, these methods require the experimental design to match precisely the modifications made to the canonical regression discontinuity design, and hence are inappropriate for our example. 

The last approach is to augment the canonical regression discontinuity design with additional covariates for which the running variable can be seen as a surrogate.
Specifically, suppose that the running variable and the potential outcome are independent conditional on some baseline covariates. 
Then, by controlling for covariates other than the running variable, the non-positivity problem associated with running variable can be circumvented.
In our illustrating example, the level of retinol acid can be regarded as such a covariate as discussed above. 
Previous studies adopting this approach include \cite{angrist2015wanna}, where they assume that these additional covariates are observed and the ``conditional independence'' assumption holds. 
However, as discussed in \cite{angrist2015wanna}, it is not always possible to find such suitable covariates. 
In the other work, \cite{eckles2020noise} introduced a noise-induced randomization framework, where an exogenous measurement error exists in the running variable while the latent true running variable is unobserved, and the conditional distribution of the running variable is pre-specified and unable to be tested. 
Without considering the identifiability of the causal effects in a larger population, they constructed a conservative bias-aware confidence interval. 

These two works follow an idea that for treated and control units to be comparable in terms of evaluating causal effects, they do not need to have the same running variable; rather, they only need to share the same auxiliary variable.
Therefore, the positivity assumption only needs to be imposed on the auxiliary variable, meaning units with any auxiliary variable value have a probability of being assigned to either the treatment or control group.
Based on this concept, it is essential to capture information on this auxiliary variable and its relation to the canonical regression discontinuity design.
However, the auxiliary variable may not be observed in a regression discontinuity design, and in such cases, information on the auxiliary variable needs to be obtained from other data sources.

Inspired by the aforementioned works and idea, we propose a novel framework that extends the canonical sharp regression discontinuity design. In our new framework,  we assume the existence of an auxiliary variable similar to the specific covariates in \cite{angrist2015wanna} or the latent true running variable in \cite{eckles2020noise}. 
Besides the main dataset used for the canonical regression discontinuity analysis, we incorporate an independent auxiliary dataset where both the running variable and the auxiliary variable are observed simultaneously. 
This framework is more practically feasible, especially when the auxiliary variable may not be directly observed in an empirical study employing a regression discontinuity design to assess treatment effects due to technological limitations or lack of awareness. 
However, it can be subsequently measured in a new prospective study. 
For example, similar frameworks have been adopted to address the measurement error issue in regression discontinuity designs \citep{davezies2017regression,bartalotti2021correction}. 

Our work has the following contributions. 
Firstly, we propose a novel data framework where an external dataset containing both the running variable and auxiliary variable is available besides the original dataset of the canonical sharp regression discontinuity design to address the extrapolation issues in regression discontinuity designs.
Secondly, we establish the identifiablity of the  average treatment effect (ATE) under this framework. 
Thirdly, we propose three estimators for ATE, show their asymptotic properties, with one of the estimators being doubly robust, and further propose  asymptotically valid inference procedure.

The rest of this paper is organized as follows. 
In Section \ref{setup}, we introduce the auxiliary variable framework for the regression discontinuity design and the data structure.
In Section \ref{identification}, we  establish the identifiability for the ATE under latent positivity and the existence of bridge function assumption. 
In Section \ref{estimation}, we  propose the estimating procedure for the target parameter.
And we show the asymptotic properties of the proposed estimators.
In Section \ref{simulation-application} conducts simulation studies to assess the finite sample performance of our methods. Besides, we apply our methods to a real world data to evaluate the average causal effect of vitamin A supplementation on autism spectrum disorders (ASD) in Chinese children.
In Section \ref{discussion}, we conclude our work and discuss some extensions.
\section{Setup}\label{setup}
\subsection{Auxiliary variable framework for regression discontinuity design}\label{framework}

Throughout this paper, we focus on the sharp regression discontinuity design.
Specifically, let $X$ be the running variable which determines the binary treatment assignment variable $W \in \{0, 1\}$, where $W = 1$ indicates receiving the treatment under evaluation, and $W=0$ otherwise.
In the sharp design, each unit receives the treatment whenever the running variable exceeds a threshold $c \in \mathbb{R}$, i.e., $W = \mathbb{I}(X \geq c)$.
Let $Y \in \mathbb{R}$ be the observed outcome of interest.
To study the causal effect of the treatment  within the regression discontinuity design, we adopt Rubin's potential outcome framework \citep{rubin1974estimating,neyman1990application}.
Here, $Y(0), Y(1)$ represent the potential outcomes given $W=0$ and $W=1$, respectively. 
And the observed outcome is $Y = WY(1) + (1 - W)Y(0)$.
we are interested in the global average of potential outcomes, defined as $\tau_w = E[Y(w)]$, where $w = 0,1$.
Additionally, the  average treatment effect is $\tau_1 - \tau_0$, that is, the  mean difference between the two potential outcomes.

\begin{figure}[h]
    \centering
    \includegraphics[scale = 0.5]{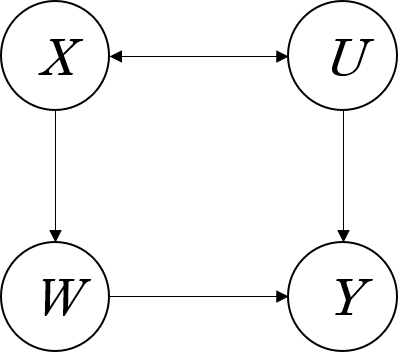}
    \caption{Graphical illustration of the regression discontinuity design with an auxiliary variable $U$. The one-way arrow indicates the causal relationship while the two-way arrow between $X$ and $U$ indicates that we do not assume a causal relationship between $X$ and $U$.}
    \label{fig:setup}
\end{figure}

Further, we assume the existence of an auxiliary variable $U$, the controlling of which can eliminate the influence of the running variable on the potential outcome. 
This assumption can be formalized as follows:
\begin{assumption}[Latent confounding]\label{latentconfounding} The running variable and treatment assignment are conditionally independent of the potential outcomes given the auxiliary variable, i.e.,\begin{align*}(X,W) \ind (Y(0), Y(1))\mid U. \end{align*}   
\end{assumption}
Assumption \ref{latentconfounding} implies that if we can observe the auxiliary variable, then the running variable can only influence the outcome through its induced treatment.
This assumption is akin to the exogeneity assumption of the noise-induced randomization framework \citep{eckles2020noise}. 
However, unlike their framework where $X$ is an observation of $U$ with measurement error, we do not assume a causal relationship between $U$ and $X$. 
It means that $U$ can also be caused by $X$. 
Moreover, Assumption \ref{latentconfounding} is more restrictive than the conditional independence assumption in \cite{angrist2015wanna}, where potential outcomes are assumed to be only mean-independent of the running variable conditional on the auxiliary variable. 
In the following, for any pair of random variables (or vectors) $A, B$ which have a joint probability density function, write $p_A(a), p_B(b)$ as their density function respectively and $p_{A|B}(a|b)$ as the conditional density of $A$ given $B$.
Figure \ref{fig:setup} provides a graphical illustration of our auxiliary variable framework for the regression discontinuity design.

\subsection{Data structure}
We propose a new data structure to evaluate the  average treatment effects in regression discontinuity design. 
Specifically, our observation includes two independent datasets: main dataset and auxiliary dataset, respectively.
The main dataset corresponds to the canonical sharp regression discontinuity design, comprising i.i.d. observations of $(X, W, Y)$ without observing $U$. 
The auxiliary sample contains the i.i.d. observations of the auxiliary variable and running variable, i.e. $(U, X)$.
We require that the distribution of $(U,X)$ in the auxiliary sample matches that of the main sample, enabling us to leverage the auxiliary variable information from the auxiliary sample.
\begin{assumption}[Exchangeability]\label{exchangeability}
Let $F_{m}(U,X)$ and $F_a(U,X)$ be the joint distributions from which the main sample and the auxiliary sample are drawn (although $U$ is unobservable in the main sample), respectively. Then, we assume $F_m(U,X) = F_a(U,X)$.
\end{assumption}
Since we have the same joint distribution of $(U,X)$ in the main sample and the auxiliary sample, throughout this paper, we let $E[\cdot]$ denote the expectation with respect to $F_m$.
We compare our data structure with the observation in the work of \cite{angrist2015wanna}, where they have i.i.d. observations of $(X, W, Y, U)$. 
In this case, the joint distribution of $(X, W, Y, U)$ can be identified, which implies this data structure is more informative.
Under  the common support assumption in \cite{angrist2015wanna}, they can estimate causal effects using traditional methods of causal inference \citep{Imbens15}. 

Moreover, our data structure is more applicable in practice. 
In a regression discontinuity design, the auxiliary variable $U$ may be ignored and not be observed together with the treatment and outcome of interest. 
For example, \citet{bartalotti2021correction} analyzed the effect of additional care received by newborns classified as very low-birth weight,   where the measured birth weight is the running variable. 
However, the measure weight is encountered with rounding error which is not captured in the main data.
In our motivating example to evaluate the effect of vitamin A supplementation on ASD disease in Chinese children \citep{Xi20}, the serum retinol concentration is the running variable and retinoic acid level is the auxiliary variable. 
But an effective technique to measure this auxiliary variable had not been established by the research team until this program finished.
These examples show that the auxiliary sample can be obtained in a subsequent research.

\section{Identification}\label{identification}
In this section, we establish nonparametric identification of the  average treatment effect.
\cite{Hahn01} have noted that in the canonical regression discontinuity design, treatment effects can only be identified at the threshold without additional assumptions, such as the constant effect assumption.
Therefore, we need to utilize information from the auxiliary dataset, i.e. the joint distribution of the auxiliary variable and the running variable.

To start with, we propose the following assumptions for identification. 
\begin{assumption}[Latent positivity]\label{positivity}
    $0 < p_{W|U}(w|U) < 1$ almost surely.
\end{assumption}

Assumption \ref{positivity} states that there are both treated and untreated units for any possible value of the auxiliary variable. 
Note that treatment assignment is determined by whether the running variable exceeds the threshold. 
Therefore, Assumption \ref{positivity} is equivalent to the possibility of observing a running variable $X$ on either side of the threshold for all possible values of the auxiliary variable $U$, which can be tested base on the auxiliary sample. 

This assumption is associated with two aspects of the joint distribution of $(U,X)$: the variability in the running variable around the auxiliary variable and the support of the auxiliary variable. 
Assumption \ref{positivity} is more likely to hold if the variation in $X$ around $U$ is large and the support of $U$ is narrow. 
In the noise-induced randomization framework \citep{eckles2020noise} where $X$ is a measure for $U$, on the one hand, the exogenous measurement error cannot be too small; in other words, this measure should not be overly precise. 
For instance, a Gaussian measurement error is sufficient to satisfy our assumption. 
On the other hand, if the support of $U$ is narrow, such as a finite-length interval, even a relatively small error, albeit finite, can result in $X$ appearing on either side of the threshold. 
Further, this assumption rules out an extreme scenario where $U$ exactly equals $X$. 
In this case, while the latent confounding assumption obviously holds, there would be no variation in the running variable based on the auxiliary variable. 
In \cite{angrist2015wanna}, the latent positivity assumption is referred to as the common support assumption, which is necessary to identify global average of potential outcomes in their framework. 

Moreover, we define the outcome bridge function  and treatment bridge function as follows, which were first introduced in proximal inference literature \citep{miao2024confounding,kallus2022causal, cui2023semiparametric,tchetgen2024introduction}. 
\begin{definition}\label{defbridgefct}
     We call $h_0 \in L^2(U,W)$ an outcome bridge function if
     \begin{align}\label{outcomebridgefct}
         E[h_0(U,W)|X, W] = E[Y\mid X, W],
     \end{align}
     and $f_0(X,W)\in L^2(X,W)$ an treatment bridge function if 
     \begin{align}\label{treatmentbridgefct}
         E[f_0(X,W)\mid U, W] = 1/p_{W|U}(W\mid U).
     \end{align}
\end{definition}
According to Definition \ref{defbridgefct}, such a function $h_0$ of $U$ and $W$ and a function $f_0$ of $X$ and $W$ serve similarly to the outcome regression function of $X$ and $W$ and the inverse propensity score function of $U$ and $W$, respectively.
Note that under Assumption \ref{latentconfounding}, the conditional expectation of $Y$ given $U$ and $W$, i.e. $E[Y\mid U,W]$ automatically satisfies the assumption of the outcome bridge function, provided it belongs to $L^2(U,W)$. 
To identify $\tau_{w}$, we impose the following assumption:
\begin{assumption}[Existence of treatment bridge function]\label{bridgefct}
    There exists a function $f_0 \in L^2(X,W)$ such that (\ref{treatmentbridgefct}) holds.
\end{assumption}
Here, we only assume the existence of the treatment bridge function but do not require their uniqueness.
With Assumption \ref{bridgefct}, we can identify the target parameter using the bridge functions.
\begin{theorem}[Identification]\label{identifiability} Under Assumption \ref{latentconfounding}-\ref{bridgefct}, the global average of the potential outcomes $\tau_w$ can be identified as follows:
\begin{align}
    \tau_w &= E[h_0(U,w)]\label{idor} \\&= E[Yf_0(X,W)\I(W=w)]\label{idps} \\&= E[Yf_0(X,W)\I(W=w) + h_0(U,w)(1-f_0(X,W)\I(W=w))]\label{iddr},
\end{align}   
where $h_0$ and $f_0$ are outcome bridge function and treatment bridge function, respectively.
\end{theorem}
Notice that the $h_0$ in \eqref{idor}  may not necessarily be the outcome regression function $E[Y\mid U,W]$.
We give examples where the treatment bridge functions exist in the supplementary material. 
We use the bridge functions for estimation in the following section.
\section{Estimation and inference}\label{estimation}
\subsection{Estimation methods}
In this subsection, we consider the estimation of the global average of the potential outcomes. 
We estimate the bridge functions by solving the conditional moment equations based on \eqref{outcomebridgefct} and \eqref{treatmentbridgefct}. 
Then, we obtain empirical estimators for the target estimators based on the identification results (\ref{idor})-(\ref{iddr}). 

We first estimate the bridge functions $f_0$ and $h_0$ based on minimax learning formulation \citep{kallus2022causal}. 
Specifically, by the definition of the bridge functions, we have the following observation.
\begin{proposition}\label{minimax}
    Let $h_0$ and $f_0$ be the outcome and treatment bridge functions, respectively. 
    Then, for any constant $\lambda> 0$, we have that
    \begin{align*}
        0 &= \sup\limits_{f' \in L^2(X,W)} E[(h_0(U,W) - Y)f'(X,W)] - \lambda ||f'||_2^2 \\ &\leq  \sup\limits_{f' \in L^2(X,W)} E[(h(U,W) - Y)f'(X,W)] - \lambda ||f'||_2^2,\\
        0 &= \sup\limits_{h' \in L^2(U,W)} E[(f_0(X,W)h'(U,W)-h'(U,0)-h'(U,1)] - \lambda ||h'||_2^2 \\ &\leq  \sup\limits_{h' \in L^2(U,W)} E[(f(X,W)h'(U,W)-h'(U,0)-h'(U,1)] - \lambda ||h'||_2^2,
    \end{align*}
    for any $h \in L^2(U,W)$ and $f \in L^2(X,W)$, where $||\cdot||$ denotes the $L^2$ norm.
\end{proposition}
Proposition \ref{minimax} introduces two functionals on which the outcome and treatment bridge functions attain their respective minimum value. 
In fact, Proposition \ref{minimax} redefines the bridge functions without considering the conditional expectation.
Based on this observation, we consider the following minimax estimator for $h_0$ and $f_0$, respectively:
\begin{align}
    \hat{h} =& \mathop{\arg\min}\limits_{h\in H} \max\limits_{f'\in F'}  \{ \hat{E}_a [h(U,W)f'(X,W)] - \hat{E}_m [f'(X,W)Y] - \lambda \norm{f'}_n^2 \}\label{minimaxh},\\
    \hat{f} =& \mathop{\arg\min}\limits_{f \in F} \max_{h' \in H'} 
    \{\hat{E}_a[f(X,W)h'(U,W) - h'(U,0)- h'(U,1)] - \lambda' \norm{h'}_a^2\}\label{minimaxf}.
\end{align}
Here, $H,H' \subset L^2(U,W), F,F'\subset L^2(X,W)$ are prespecified function spaces; $\lambda, \lambda' > 0$ are prespecified constants; $\hat{E}_m, \hat{E}_a$ and $\hat{E}_n$ denote taking empirical expectation over the main sample, over auxiliary sample and over two sample sets combined together, respectively; $\norm{f'}_n^2 = \hat{E}_n [f'(X,W)^2]$ and $\norm{h'}_a^2=\hat{E}_a[h'(U,W)^2]$ denotes the empirical $L^2$ norms with respect to the whole sample and the auxiliary sample, respectively; $\hat\pi_w(u)$ is the estimator for the propensity score $p_{W\mid U}(w\mid u)$. 

Then, we propose three estimators for $\tau_w$:
\begin{align}
    &\hat\tau_w^h = \hat{E}_a[\hat{h}(U,w)],\\
    &\hat\tau_w^f = \hat{E}_m[\hat{f}(X,W)\I(W=w)Y], \\
    &\hat\tau_w^\text{dr} = \hat{E}_m [\hat{f}(X,W)\I(W=w)Y] + \hat{E}_a[(1-\hat{f}(X,W)\I(W=w))\hat{h}(U,w))].
\end{align}
These three estimators correspond to the three identification formulae (\ref{idor})-(\ref{iddr}), respectively, paralleling the causal effects' estimators in the classical causal inference: the outcome regression estimator , the inverse propensity  weighted estimator and the doubly robust estimator \citep{bang2005doubly,cui2023semiparametric}.
The estimator $\hat\tau_w^h$ only depends on the estimated outcome bridge function; the estimator $\hat\tau_w^f$ only depends on the estimated treatment bridge function;  and the doubly robust estimator $\hat\tau_w^\text{dr}$ depends on both bridge functions.
Note that, for the terms involving the auxiliary variable $U$, we can only take an empirical expectation over the auxiliary sample, while for the terms involving the outcome variable $Y$, we can only take an empirical expectation over the main sample.

\subsection{Asymptotic properties} \label{asymptoticproperties}
In this subsection, we study the asymptotic properties of the three proposed estimators for the target causal parameters. 
We first show the convergence of the minimax estimators for the bridge function under  assumptions in terms of projected mean-squared errors. 
Then, we show that the estimators for the target parameters are consistent. 
Finally, we establish the asymptotic normality of the doubly robust estimator. 

We establish the convergence rate of the estimators for the bridge functions. 
Note that, we define and estimate the bridge functions by leveraging the conditional expectation equations, which implies that all the information of the estimated bridge functions can be obtained from the perspective of a conditional expectation, i.e. a projection. 
Therefore, we study the convergence of the bridge functions based on the projected mean-squared error:
\begin{align*}
    ||E[\hat{h}(U,W) - h_0(U,W)\mid X,W]||_2^2,\\
    ||E[\hat{f}(X,W) - f_0(X,W)\mid U,W]||_2^2.
\end{align*}
Besides we introduce a function class complexity measure called critical radius \citep{bartlett2005local}.
\begin{definition}[Localized Rademacher complexity and critical radius]
    Given a class $\mathcal{G}$ of functions of variables $O$, its empirical localized Rademacher complexity with respect to the sample $\{O_1,\dots,O_n\}$ and radius $\eta > 0$ is defined as 
    \begin{align*}
        \mathcal{R}_n(\eta;\mathcal{G}) = \frac{1}{2^n} \sum_{\epsilon\in \{-1,+1\}^n} \left\{\sup\limits_{g \in \mathcal{R}:E_n[g^2]\leq \eta^2}\frac1n \sum_i \epsilon_i g(O_i) \right\}.
    \end{align*}
    When $\eta = +\infty$, the quantity $\mathcal{R}_n(+\infty;\mathcal{G})$ is called the (global) Rademacher complexity. The critical radius of $\mathcal{G}$ is the smallest positive $\eta$ that satisfies the inequality: $\mathcal{R}_n(\eta;\mathcal{G}) \leq \eta^2/||\mathcal{G}||_\infty$, where $||\mathcal{G}||_\infty = \sup\limits_{g\in\mathcal{G}}\{||g||_\infty\}$ denotes the sup norm of $\mathcal{G}$.
\end{definition}
Further, we introduce a strict latent positivity assumption which is a more restricted form of Assumption \ref{positivity}.
\begin{assumption}[Strict latent positivity]\label{strictpositivity}
    For $w = 0,1$, we assume $\varepsilon < p_{W|U}(w|U) < 1-\varepsilon$ almost surely for some $\varepsilon > 0$.
\end{assumption}

Let the sample size of the main sample and the auxiliary sample be $n_1$,$n_2$, respectively, and further $n = n_1+n_2$ be the total sample size. For a set $S$ of a linear space, we call $S$ symmetric if $-s \in S$ for any $s \in S$; we call $S$ star-shaped (around the origin) if $as \in S$ for all $a \in [0,1]$ and any $s \in S$. In the following result, we establish the convergence rate of the outcome bridge functions. 
\begin{theorem}[Convergence rate of $\hat{h}$]\label{asymptotich}
    Assume that $||H||_\infty, ||F'||_\infty, |Y|$ are bounded, there is an outcome bridge function $h_0\in H$, and further assume the following conditions:\\
    \quad 1. $F'$ is symmetric and star-shaped.\\
    \quad 2. Let $\eta_{h,n_1}, \eta_{h,n_2}, \eta_{h,n}$ upper bounds the critical radii of $F'$ and $\mathcal{G}_{h_0}$
    with respect to the main sample, the auxiliary sample and the sample combined, where the class $\mathcal{G}_{h_0}$ is defined as\begin{align*}\mathcal{G}_{h_0} = \{(h-h_0)f: h\in H, f \in F'\}.\end{align*}
    Then, $\eta_{h,n} \to 0$ in probability and $\max(\eta_{h,n_1},\eta_{h,n_2}) = O(\eta_{h,n})$.\\
    3. For any $h\in H$, we have $E[h(U,W) - h_0(U,W)|X,W] \in F'$.
    
    Then under Assumptions \ref{latentconfounding}-\ref{strictpositivity}, we have with probability $1-\tilde\delta$,
    \begin{align*}
        \norm{E[\hat{h}(U,W) - h_0(U,W)\mid X,W]}_2 = O\left(\eta_{h,n} + \sqrt{(1+\log(1/\delta))/n}\right),
    \end{align*}
    where $\tilde\delta = c_0\exp(-c_1n\eta_{n}^2/\norm{F'}_\infty^2)+\delta$, $\eta_n = \eta_{h,n}+\sqrt{c_2\log(c_3/\delta)/n}$, and $c_0, c_1, c_2, c_3$ are positive constants.
\end{theorem}
In Theorem \ref{asymptotich}, we assume that the selected function spaces need to be sufficiently rich so that the $H$ should contain the true outcome bridge function and that $F'$ should cover $E[h-h_0\mid X,W]$ for all $h \in H$. 
Theorem \ref{asymptotich} states that the convergence rate of the estimated outcome bridge function is determined by the critical radii of $F'$ and $\mathcal{G}_{h_0}$.
It is worth noting that the term $\sqrt{c_2\log(c_3/\delta)/n}$ in $\eta_n$ makes the term $c_0\exp(-c_1n\eta_{n}^2/\norm{F'}_\infty^2)$ in $\tilde\delta$ at least a polynomial order of $\delta$.
Thus, as $\delta$ goes to 0, $1-\tilde\delta$ is a high probability converging to 1.
There is a similar phenomenon in Theorem \ref{asymptoticf}.

Then we establish the consistency of the estimator  $\hat\tau_w^h$  for the target parameter.
\begin{theorem}[Consistency of $\hat\tau_w^h$]\label{asymptotictauh}
    Assume that $\{h(U,w)\mid h \in H\}$ is a Glivenko-Cantelli class. Then, under the conditions in Theorem \ref{asymptotich}, we have $\hat\tau_w^h$ is consistent. 
\end{theorem}

Similarly, we establish the asymptotic properties for the estimated treatment bridge function $\hat{f}$ and the estimator for the target parameter $\hat\tau_w^f$.
\begin{theorem}[Convergence rate of $\hat{f}$]\label{asymptoticf}
    Assume that $||F||_\infty, ||H'||_\infty$ are bounded, there is a treatment bridge function $f_0\in F$, and further assume the following conditions:\\
    \quad 1. $H'$ is symmetric and star-shaped.\\
    \quad 2. Let $\eta_{f,n_2}$ upper bounds the critical radius of $H'$ and $\mathcal{G}_{f_0}$
    with respect to the auxiliary sample and $\eta_{f,n_2} \to 0$ in probability, where the class $\mathcal{G}_{f_0}$ is defined as\begin{align*}\mathcal{G}_{f_0} = \{(f-f_0)h: f\in F, h \in H'\}.\end{align*}
    3. For any $f\in F$, we have $E[f(U,W) - f_0(U,W)|X,W] \in H'$.
    
    Then under Assumptions \ref{latentconfounding}-\ref{strictpositivity}, we have with probability $1-\tilde\delta$,
    \begin{align*}
        \norm{E[\hat{f}(X,W) - 1/\pi_W(U)\mid U,W]}_2 \leq O(\eta_{h,n_2} + \sqrt{(1+\log(1/\delta))/n_2}).
    \end{align*}
    where $\tilde\delta = c_0 \exp(-c_1 n_2 \eta_{n_2}^2/\norm{H'}_\infty^2)+\delta$, $\eta_{n_2} = \eta_{f,n_2} + \sqrt{c_2\log(c_3/\delta)/n_2}$, and $c_0, c_1, c_2, c_3$ are positive constants.
\end{theorem}
\begin{theorem}[Consistency of $\hat\tau_w^f$]\label{asymptotictauf}
    Assume that $F$ is a Glivenko-Cantelli class.
    Then, under the conditions in Theorem \ref{asymptoticf}, we have $\hat\tau_w^h$ is consistent. 
\end{theorem}

Moreover, we derive the asymptotic property for the doubly robust estimator $\hat\tau_w^\text{dr}$ with double robustness. 
We first define the ill-posedness measures as followed:
\begin{align*}
    c^F =& \sup\limits_{f\in F }\inf\left\{ \frac{||f(X,W)-f_0(X,W)||_2}{||E[f(X,W)-f_0(X,W)|U,W]||_2}\mid f_0\in F, E[f_0(X,W)|U, W] = 1/p_{W|U}(W|U)\right\};\\
    c^H =& \sup\limits_{h\in H}\inf\left\{ \frac{\norm{h(U,W)-h_0(U,W)}_2}{\norm{E[h(U,W)-h_0(U,W)\mid X,W]}_2}\mid h_0 \in H,E[h_0(U,W)\mid X,W]=E[Y\mid X,W] \right\}.
\end{align*}
Note that our definition of ill-posedness measures is different from \cite{kallus2022causal} in that their definition of ill-posedness measures depends on the choice of bridge functions, while our definition does not require specifying fixed bridge functions.
\begin{theorem}[Asymptotic decomposition for $\hat\tau^\text{dr}$]\label{asymptoticdb} Assume that $\rho_n=n_1/n$ converges to a constant $\rho\in (0,1)$ in probability,  $F$ and $\{h(U,w):h\in H\}$ are Donsker classes, and  Assumptions \ref{latentconfounding}-\ref{strictpositivity} hold. Then we have:
\begin{itemize}
    \item[(1)] The doubly robust estimator $\hat\tau_w^\text{dr}$ is consistent if either $||E[\hat{h}(U,W)- Y|X,W]||_2$ or $||E[\hat{f}(X, W) - f_0(X, W)| U,W]||_2$ converges to $0$ in $L_2$. Further assuming both $||E[\hat{h}(U,W)- Y|X,W]||_2$ and $||E[\hat{f}(X, W) - f_0(X, W)| U,W]||_2$ converge to $0$ in $L_2$, we have that:
    \begin{align*}
        \hat\tau_w^\text{dr} - \tau_w =O_P\left(\min\{c^H,c^F\} ||E[\hat{h}(U,W)-Y|X,W]||_2\cdot||E[\hat{f}(X,W)-f_0(X,W)|U,W]||_2+n^{-1/2}\right). 
    \end{align*}
    \item[(2)]     Moreover, assume there is a unique treatment bridge function $f_0 \in F$ and a unique outcome bridge function $h_0 \in H$ and $c^F c^H < \infty$, and there exist $C, t \geq 0$, such that  $\max\{N(F, \norm{\cdot}_2,s),$ $ N(H, \norm{\cdot}_2, s)\}\leq Cs^{-t}$, where $N(\Theta, \norm{\cdot}_2,s)$ denotes the covering number of the space $\Theta$ with respect to $L^2$ norm and radius $s$. 
    Then, we have the following decomposition:
    \begin{align*}
        \hat\tau_w^\text{dr} - \tau_w =&(\hat{E}_m - E) [\I(W=w)f_0(X,W)Y] + (\hat{E}_a - E) [(1-\I(W=w)f_0(X,W))h_0(U,w)] \\&+ O_P(\min\{c^H,c^F\} ||E[\hat{h}(U,W)-Y|X,W]||\cdot||E[\hat{f}(X,W)-f_0(X,W)|U,W]||) + o_P(n^{-1/2}).
    \end{align*}
\end{itemize}
\end{theorem}
Theorem \ref{asymptoticdb} states that if $\max\{c^F,c^H\} ||E[\hat{h}(U,W)-Y|X,W]||\cdot||E[\hat{f}(X,W)-f(X,W)|U,W]|| = o_p(n^{-1/2})$, the doubly robust estimator is asymptotically normal, with the asymptotic variance term:
\begin{align*}
    \sqrt{\frac1{\rho} S_m^2 + \frac1{1-\rho} S_a^2},
\end{align*}
where $S_m^2$ is the variance of $\I(W=w)f_0(X,W)Y$ and $S_a^2$ is the variance of $(1-\I(W=w)f_0(X,W))h_0(U,w)$.
The convergence rates of estimated bridge functions in terms of their projected mean-squared error depend on the critical radii of $F'$ and $H'$, which were discussed in widely-used function spaces in \cite{kallus2022causal}.
Moreover, Theorem \ref{asymptoticdb} implies that $\hat\tau_w^\text{dr}$ is doubly robust in the sense that it is consistent if either $\hat{h}$ or $\hat{f}$ converges in terms of the projected mean squared error.
\section{Simulations and real data analysis}\label{simulation-application}
\subsection{Simulations}\label{simulation}
In this subsection, we conduct simulations to assess the finite sample performance of our methods.
The results show the validity of the three estimators, $\hat\tau_w^h, \hat\tau_w^f, \hat\tau_w^\text{db}$ and their induced confidence intervals in the simulated datasets.
We report the mean squared error (MSE) of the estimators to illustrate the estimation accuracy.
For each estimator, we construct the bootstrap confidence intervals.
For these confidence intervals, we report their coverage rates and average lengths.

\begin{table}[htp]
    \centering
    \begin{tabular}{r|ccc}
    \toprule
         & auxiliary variable $(U)$ & Running variable $(X\mid U)$& Outcome $(Y\mid U,X,W)$ \\ \midrule
        Setting 1 & $Bernulli(0.5)$ & $\mathcal{N}(U-0.5,1)$  & $\mathcal{N}(U+2W,1)$ \\ 
        Setting 2 & $Uniform(0,1)$& $\mathcal{N}(U-0.5,1)$ &  $Bernulli(1/(1+\exp(0.6U-2W)))$ \\ 
    \bottomrule
    \end{tabular}
    \caption{Setups of data generation.}
    \label{tab:datageneration}
\end{table}
In our simulation studies, we consider two setups of data generation as shown in Table \ref{tab:datageneration}.
In the first setup, we consider a binary auxiliary variable and a continuous outcome, while in the second setup, we consider a continuous auxiliary variable and a binary outcome. 
In each setup, the running variable follows normal distribution and the threshold $c$ is fixed to be 0, which means that each unit receives a treatment if the running variable exceeds $0$. 
We consider the sample sizes of the main sample and the auxiliary sample as $(n_1,n_2) = (100 \times k, 100 \times k), k =1,2,5,10$.


We now elaborate on how to solve the optimization problems (\ref{minimaxh}) and (\ref{minimaxf}) to obtain $\hat{h}$ and $\hat{f}$. 
Since they are both minimax optimization problems, we first seek closed form solutions of the inner optimization problems and then solve the outer problems by gradient descent. 
In particular, for the inner problems, we specify the classes $F'$ and $H'$ as bounded linear classes: 
\begin{align*}
    &F' = \{(x,w) \mapsto \alpha_{1}^T \phi(x,w): \alpha_{1} \in \mathbb{R}^{d_1},||\alpha_{1}|| \leq c_1 \}, \\
    &H' = \{(u,w)\mapsto \alpha_{2}^T \psi(u,w): \alpha_{2} \in \mathbb{R}^{d_2}, ||\alpha_{2}|| \leq c_2 \},
\end{align*}
given basis functions $\phi(x,w), \psi(u,w)$ of dimensions $d_1,d_2$. Similar to \cite{kallus2022causal}'s estimating procedures, we add a penalty term on the objective function as we constrain the norms of $\alpha_1,\alpha_2$: 
\begin{align*}
    &\hat{h} = \mathop{\arg\min}\limits_{h\in H} \max\limits_{f'\in F'}  \{ \hat{E}_a [h(U,W)f'(X,W)] - \hat{E}_m [f'(X,W)Y] - \lambda \hat{E}_n [(f'(X,W))^2] - \gamma_1||\alpha_{1}||^2\},\\
    &\hat{f} = \mathop{\arg\min}\limits_{f \in F} \max_{h' \in H'} 
    \{\hat{E}_a[f(X,W) h'(U,W) - h'(U,0) - h'(U,1)] - \lambda'\hat{E}_a[(h'(U,W))^2] - \gamma_2 ||\alpha_{2}||^2\}.
\end{align*}
Then we have
\begin{align*}
    \hat{h} = \mathop{\arg\min}\limits_{h\in H}& (\hat{E}_a[\phi h] - \hat{E}_m[\phi Y] )^T (\lambda \hat{E}_n[\phi \phi^T] + \gamma_1 I_{d_1})^{-1}(\hat{E}_a[\phi h] - \hat{E}_m[\phi Y] ),\\
    \hat{f} = \mathop{\arg\min}\limits_{f \in F}& \hat{E}_a[f \psi-\psi_0-\psi_1]^T (\lambda' \hat{E}_a[\psi\psi^T] + \gamma_2 I_{d_2})^{-1} \hat{E}_a[f\psi - \psi_0-\psi_1],
\end{align*}
with $\psi_1 = \psi(u,1)$ and $\psi_0 = \psi(u,0)$.
For the outer optimization problems, we specify the function $h$ and $f$ as two-layer neural networks. We optimize them by gradient descent.

For each estimator, we construct the 95\% bootstrap confidence interval by resampling the dataset 1000 times.
We calculate MSE for each estimator and coverage rate and average length for each confidence interval by Monte Carlo method to verify the asymptotic validity of the estimating and inference procedure.
We compare the performance of different estimators $\hat\tau_w^f,\hat\tau_w^h,\hat\tau_w^\text{db}$ when one of the bridge functions is misspecified.

Tables \ref{tab:setting_15} and \ref{tab:setting_2} shows the MSE of the three estimators under Setting 1 and Setting 2, respectively. 
The MSE decreased towards zero as sample size increases, showing the consistency of our estimator.
And the results provide strong evidence that, under correct specification of the bridge functions, our proposed estimator performs well in both small and large samples. 
Moreover, Tables \ref{tab:setting_1_interval} and \ref{tab:setting_2_interval} show the performance of the bootstrap confidence interval.
The bootstrap confidence intervals generally maintain an adequate coverage rate close to the nominal level (95\%) across different scenarios.
The bootstrap confidence intervals perform better with larger sample sizes, showing more stable and consistent coverage rates. 

\begin{figure}[h]
    \centering
    \includegraphics[scale = 0.75]{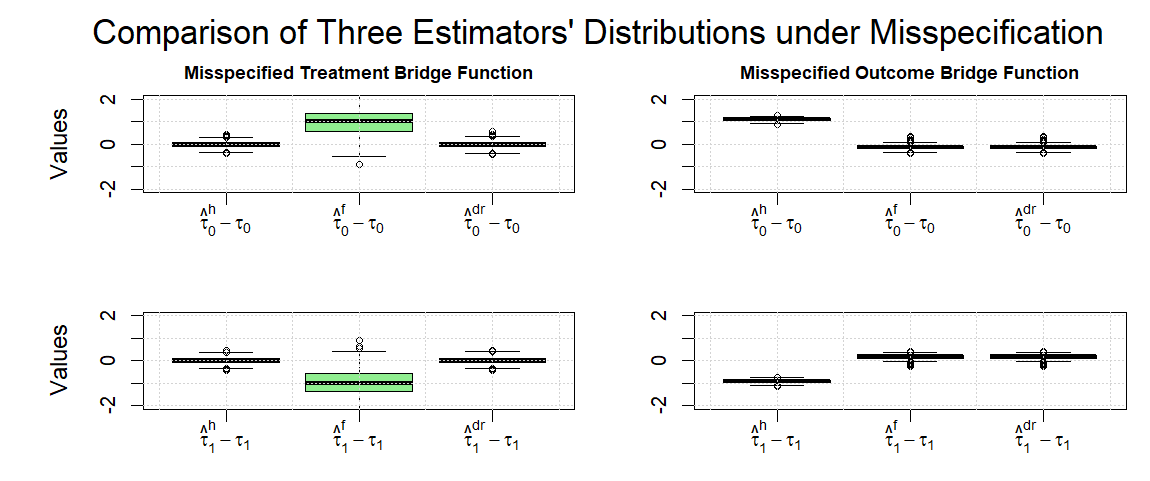}
    \caption{Performance of the three estimating procedures under setting 1 when one of the bridge functions is incorrectly specified.}
    \label{fig:sim_dr1}
\end{figure}

\begin{figure}[h]
    \centering
    \includegraphics[scale = 0.75]{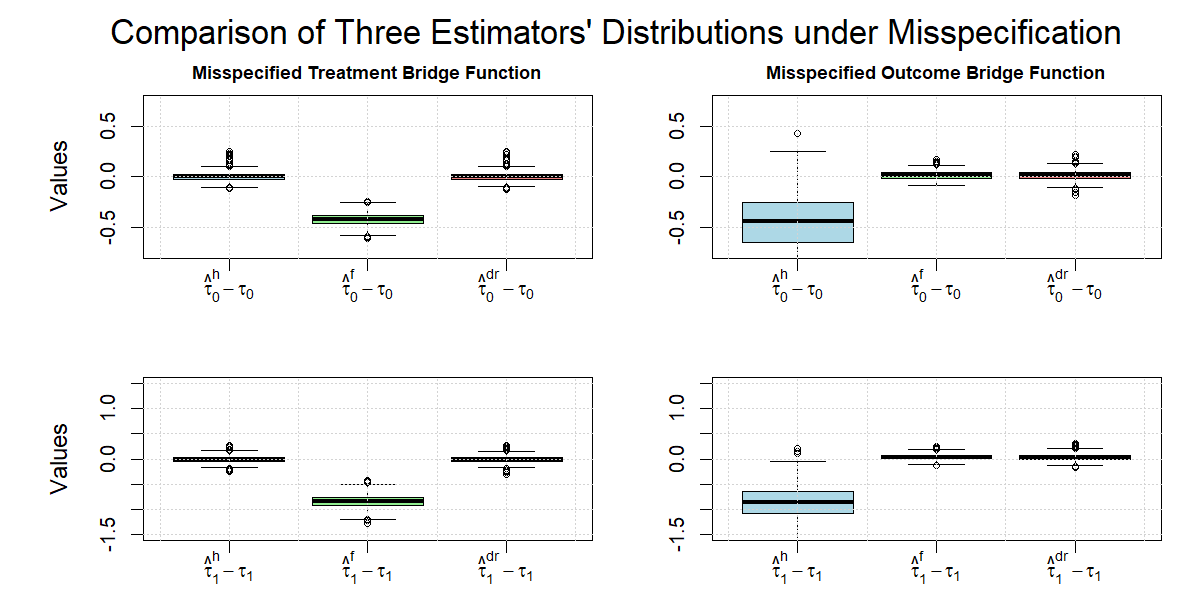}
    \caption{Performance of the three estimating procedures under setting 2 when one of the bridge functions is incorrectly specified.}
    \label{fig:sim_dr2}
\end{figure}

Besides, when both the treatment and outcome bridge functions are correctly specified, we generally observe an equally good performance for these estimators and the confidence intervals.
Based on our results, the bootstrap confidence interval for the outcome regression estimator performs more unstable, in terms of the relatively low coverage rates in small sample sizes.
However, when one of the bridge functions is incorrectly specified (Figures \ref{fig:sim_dr1} and \ref{fig:sim_dr2}), the performance of the estimators $\hat\tau_w^h$ and $\hat\tau_w^f$ deteriorates significantly, while the doubly robust estimator $\hat\tau_w^\text{db}$ remains consistent.
We present the details of the misspecified model in the supplementary material.
Therefore, the doubly robust estimator and its bootstrap confidence intervals are generally the most reliable choice, as they offer protection against potential misspecifications of the bridge functions. 
In practice, we recommend to use the doubly robust estimator when there is uncertainty about the correct modeling of the bridge functions.
Additionally,  bootstrap confidence interval for the doubly robust estimator can provide more robust inference.

\begin{table}[htp]
    \centering
    \begin{tabular}{r|ccc}
    \toprule
        number of data & MSE of $\hat\tau_w^h$ & MSE of $\hat\tau_w^f$ & MSE of $\hat\tau_w^\text{dr}$ \\ \midrule
        100 & (0.1307, 0.1427) & (0.1015, 0.1174) & (0.1929, 0.2005) \\  
        200 & (0.0791, 0.0838) & (0.0629, 0.0671) & (0.1291, 0.1260) \\  
        500 & (0.0405, 0.0395) & (0.0381, 0.0443) & (0.0663, 0.0631) \\  
        1000 & (0.0188, 0.0186) & (0.0303, 0.0352) & (0.0312, 0.0282) \\  
    \bottomrule
    \end{tabular}
    \caption{Mean squared errors (MSEs) for the three estimators under the setting 1.}
    \label{tab:setting_15}
\end{table}

\begin{table}[htp]
    \centering
    \resizebox{\linewidth}{!}{
    \begin{tabular}{r|cccccc}
    \toprule
         data size ($w$) & coverage of $\hat\tau_w^h$ & length of $\hat\tau_w^h$ & coverage of $\hat\tau_w^f$ & length of $\hat\tau_w^f$ & coverage of $\hat\tau_w^\text{dr}$ & length of $\hat\tau_w^\text{dr}$ \\ \midrule
        100 (0) & 92.3\% & 1.5597 & 97.9\% & 2.3508 & 98.5\% & 2.4633 \\  
        100 (1) & 92.2\% & 1.5671 & 94.7\% & 2.3177 & 96.5\% & 2.4123 \\  \midrule
        200 (0) & 96.1\% & 1.1688 & 96.7\% & 1.3632 & 97.0\% & 1.4160 \\  
        200 (1) & 96.8\% & 1.1796 & 94.5\% & 1.2547 & 96.1\% & 1.3000 \\  \midrule
        500 (0) & 97.0\% & 0.7881 & 94.3\% & 0.8320 & 96.2\% & 1.0268 \\  
        500 (1) & 97.3\% & 0.7989 & 91.3\% & 0.7493 & 95.7\% & 0.9830 \\  \midrule
        1000 (0) & 96.7\% & 0.5681 & 93.0\% & 0.6148 & 96.7\% & 0.7388 \\  
        1000 (1) & 97.0\% & 0.5702 & 91.9\% & 0.5935 & 97.7\% & 0.7177 \\  
    \bottomrule
    \end{tabular}
    }
    \caption{Coverage rate and average length of the 95\% bootstrap confidence interval based on the three estimators under the setting 1.}
    \label{tab:setting_1_interval}
\end{table}

\begin{table}[htp]
    \centering
    \begin{tabular}{r|ccc}
    \toprule
        number of data & MSE of $\hat\tau_w^h$ & MSE of $\hat\tau_w^f$ & MSE of $\hat\tau_w^\text{dr}$ \\ \midrule
        100 & (0.0124, 0.0071) & (0.0067, 0.0038) & (0.0097, 0.0054) \\ 
        200 & (0.0057, 0.0038) & (0.0035, 0.0019) & (0.0047, 0.0031) \\ 
        500 & (0.0025, 0.0016) & (0.0014, 0.0007) & (0.0021, 0.0013) \\ 
        1000 & (0.0013, 0.0008) & (0.0008, 0.0004) & (0.0012, 0.0007) \\ 
    \bottomrule
    \end{tabular}
    \caption{Mean squared errors (MSEs) for the three estimators under the setting 2.}
    \label{tab:setting_2}
\end{table}

\begin{table}[htp]
    \centering
    \resizebox{\linewidth}{!}{
    \begin{tabular}{r|cccccc}
    \toprule
        data size ($w$) & coverage of $\hat\tau_w^h$ & length of $\hat\tau_w^h$ & coverage of $\hat\tau_w^f$ & length of $\hat\tau_w^f$ & coverage of $\hat\tau_w^\text{dr}$ & length of $\hat\tau_w^\text{dr}$ \\ \midrule
        100 (0) & 91.8\% & 0.4253 & 96.5\% & 0.4089 & 97.0\% & 0.4352 \\ 
        100 (1) & 91.1\% & 0.3394 & 96.5\% & 0.3203 & 97.2\% & 0.3405 \\   \midrule
        200 (0) & 93.4\% & 0.2819 & 94.9\% & 0.2373 & 95.6\% & 0.2483 \\  
        200 (1) & 93.6\% & 0.2270 & 95.0\% & 0.1826 & 95.0\% & 0.1914 \\   \midrule
        500 (0) & 95.6\% & 0.1883 & 94.4\% & 0.1446 & 97.3\% & 0.1941 \\  
        500 (1) & 95.5\% & 0.1524 & 95.5\% & 0.1109 & 97.5\% & 0.1544 \\  \midrule
        1000 (0) & 95.1\% & 0.1339 & 92.7\% & 0.0982 & 97.1\% & 0.1379 \\  
        1000 (1) & 94.9\% & 0.1085 & 93.6\% & 0.0773 & 97.0\% & 0.1110 \\  
    \bottomrule
    \end{tabular}
    }
    \caption{Coverage rate and average length of the 95\% bootstrap confidence interval based on the three estimators under the setting 2.}
    \label{tab:setting_2_interval}
\end{table}

\newpage
\subsection{Real data application}\label{application}
In this subsection, we apply our method  to the motivating example mentioned in Section \ref{introduction}. The medical question is whether taking vitamin A impacts the reduction and recovery of autism spectrum disorders in children. 
Autism spectrum disorder (ASD) is a term used to describe a constellation of early-appearing social communication deficits and repetitive sensory–motor behaviors associated with a strong genetic component as well as other causes \citep{Lord2018}.
Children with ASD are susceptible to multiple comorbidity, which affects the rehabilitation and quality of life \citep{Lord2018}.
Vitamin A is essential for the brain development, which is transported in the blood in the form of retinol, and functions as retinoic acid within tissues \citep{Shearer2012}.
While vitamin A interventions have been widely employed to reduce blindness and mortality within preschool children, evidence regarding their causal effects on  alleviating the severity of autism spectrum disorders remains limited.   

Our analysis is based on the ASD data collected from the vitamin A supplementary program, with the aim of reducing the severity of autism spectrum disorders in children \citep{Xi20}.
The data covers the children aged from 3 to 8 years who had autism spectrum disorders, with a total observations of 149, with a canonical sharp regression discontinuity design. 
The running variable is the baseline serum retinol concentration, a widely used measurement for the level of vitamin A.
If the serum retinol concentration is lower than 1.05 $\mu$mol/L, the participant is assigned to adjuvant vitamin A therapy, that is, the treatment group; otherwise, he/she is assigned to behavioral therapy, that is, the control group.
The outcome is the Social Reaction Scale (SRS) score at 6 months, which is used to evaluate the severity of ASD.
Previous analysis on the ASD data did not evaluate the  average treatment effect of the program for the whole children population.
\cite{Xi20} evaluated the correlation of the outcomes between the treated and controlled groups, where their results did not have causal explanation.
\cite{Feng2024causal} evaluated the local treatment effect within the subpopulation at the threshold of the running variable, without answering the  average treatment effect.

\begin{figure}[h]
    \centering
    \includegraphics[width=0.8\linewidth]{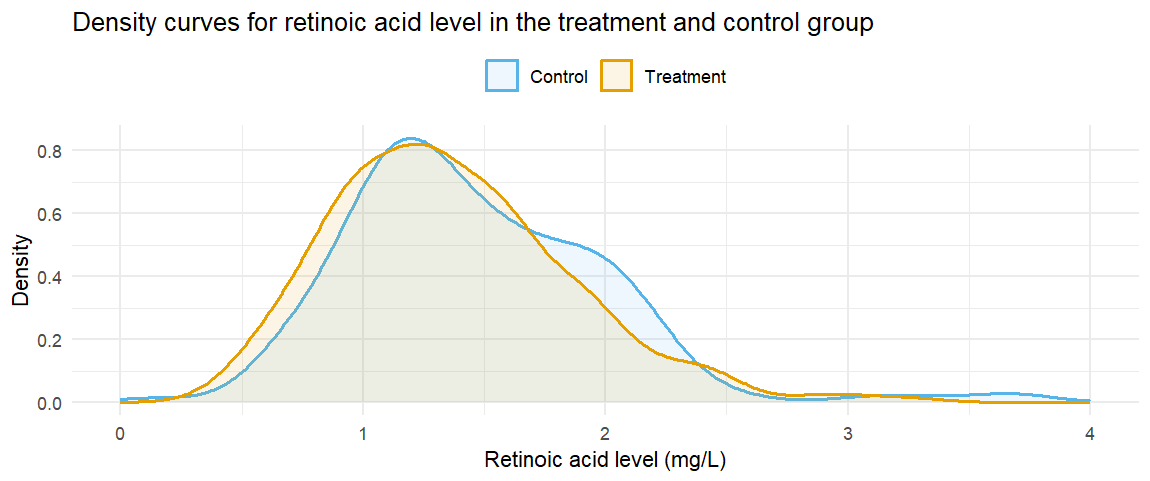}
    \caption{Histograms of retinoic acid for two subgroups defined by whether the serum retinol concentration is lower than 1.05$\mu$mol/L in the auxiliary dataset.}
    \label{fig:hist_realdata}
\end{figure}

Based on biological knowledge, serum retinol, the running variable $X$ needs to be transformed into retinoic acid to function in tissues \citep{Shearer2012}.
Hence, we think that the serum retinol concentration and the outcome (SRS score) are likely to be independent conditional on retinoic acid level, that is, we consider retinoic acid as the auxiliary variable $U$, and Assumption \ref{latentconfounding} holds in this scenario. 
Recently, \cite{Xi20}'s team has proposed a novel technique for measuring retinoic acid level and obtained an independent dataset of children with autism spectrum disorders in which the serum retinol concentration and retinoic acid level were simultaneously measured \citep{Feng2024association}, with a total number of 378.
We justify Assumption \ref{positivity} based on the auxiliary dataset with two histograms of retinoic acid for two subgroups defined by whether the serum retinol concentration is lower than 1.05$\mu$mol/L (Figure \ref{fig:hist_realdata}).
With this auxiliary dataset, we first evaluate the vitamin A supplementation on reducing the severity of the autism spectrum disorders over whole population for the main dataset.

Following the same estimation and inference procedure as the setting 2 in Section \ref{simulation}, we obtain three estimates of the average SRS score had children with autism spectrum disorders received the vitamin A adjuvant therapy or merely behavioral therapy, respectively, alongside the corresponding 95\% confidence intervals. 
Moreover, we calculate the 95\% bootstrap confidence interval of $\tau_1 - \tau_0$, the average effects of vitamin A supplement, based on the doubly robust estimator, which is (-12.48, 0.93). 
Our results numerically demonstrates a positive effect of vitamin A supplement on alleviating the severity of autism spectrum disorders.
This is consistent with the conclusions drawn from the results of association analysis \citep{Xi20} and local average treatment effect\citep{Feng2024causal}.
However, due to the lack of statistical significance, further evidence is required to enhance the power of our conclusion.

\begin{table}[h]
    \centering
    \begin{tabular}{r|cc}
    \toprule
        Estimator & Treatment  & Control  \\ \midrule
        $\hat\tau_w^h$ & 90.26 (85.62, 96.61) & 97.71 (92.42, 103.21) \\ 
        $\hat\tau_w^f$ & 91.54 (86.64, 96.87) & 97.06 (92.89, 101.81) \\  
        $\hat\tau_w^\text{dr}$ & 90.87 (86.61, 96.92) & 98.02 (92.95, 101.89) \\ 
    \bottomrule
    \end{tabular}
    \caption{Estimating results of average potential outcome (SRS score) had children with autism spectrum disorders received the vitamin A adjuvant therapy or merely behavioral therapy, respectively, alongside the corresponding 95\% confidence intervals}
    \label{tab:application_interval}
\end{table}

\section{Discussion}\label{discussion}
In this study, we propose a novel framework and statistical methodology to enhance the extrapolation capabilities of regression discontinuity designs, which is an essential challenge in this research area. 
The key idea is an introduction of a latent  variable such that a non-positivity problem can thus be transformed into an unmeasured confounding problem.
In previous works adopting this idea, \cite{angrist2015wanna} assumed the observation of the auxiliary variable within the regression discontinuity design and derive the identification of average treatment effect; \cite{eckles2020noise} assumed no further observation based on the canonical regression discontinuity design and gives conservative confidence intervals without identifying the average treatment effect.
Compared with them, we assume an extra independent dataset consisting of running variable and auxiliary variable, which is weaker than \cite{angrist2015wanna} while still identifying the parameter of interest.

Our study provides a practical guideline for researchers to evaluate treatment or policy effects in the situation where only regression discontinuity designs are available.
Specifically, researchers are supposed to identify another factor which serves as the auxiliary variable $U$, then draw an independent sample of the running variable and auxiliary variable from the target population, and finally analyze the data using our proposed procedure.

Building on our findings, future research should explore several avenues. 
Firstly, our framework allows for the evaluation of causal effects across a wider range of target populations. 
For instance, in our motivating example, we can assess the covariate-adjusted average treatment effect of vitamin A supplement on autism spectrum disorders conditional on the baseline vitamin A level. 
This can help find a reference point for vitamin A deficiency in the context of treating autism spectrum disorders.
Secondly, since regression discontinuity designs are a special case of observations without the positive assumption \citep{Imbens15}, we can extend our method to address non-positivity issues in general observational data.
\newpage
\bibliographystyle{apalike}
\bibliography{reference}
\newpage
\section{Appendix}
\subsection{Proof of Theorem \ref{identifiability}}
Note that we have
\begin{align*}
    E[h_0(U,w)] =& E[h_0(U,W)\frac{\I(W=w)}{\pi_w(U)}] \\
    =& E[h_0(U,W)\I(W=w)E[f_0(X,W)\mid U,W]] \tag{Definition of treatment bridge function}\\
    =& E[h_0(U,W)\I(W=w)f_0(X,W)]\\
    =& E[E[h_0(U,W)\mid X,W]\I(W=w)f_0(X,W)]\\
    =& E[Y\I(W=w)f_0(X,W)] \tag{Definition of outcome bridge function}\\
    =& E[Y(w)\I(W=w)f_0(X,W)].
\end{align*} 
We obtain that the identification formulae (\ref{idor})-(\ref{iddr}) are equal. 
Then, it suffices to show any one of them equals the target parameter $\tau_w$:
\begin{align*}
    E[Y(w)\I(W=w)f_0(X,W)] =& E[E[Y(w)\I(W=w)f_0(X,W)\mid Y(w),U]] \\
    =& E[Y(w)E[\I(W=w)f_0(X,W)\mid U]]\tag{Latent confounding}\\
    =&E[Y(w)E[\I(W=w)E[f_0(X,W)\mid U,W]\mid U]] \\
    =& E[Y(w)E[\frac{\I(W=w)}{\pi_W(U)}\mid U]] \tag{Definition of treatment bridge function} \\
    =&E[Y(w)]
\end{align*}
\begin{remark}
    Note that to establish $E[h_0(U,w)] = \tau_w$, we first show that $E[h_0(U,w)] = E[h_0(U,W)\I(W=w)f_0(X,W)]$, where we utilize the existence of the treatment bridge function.  
\end{remark}

\subsection{Proof of Proposition \ref{minimax}}
We focus on the first inequality.
Note that, by Hölder's inequality and AM-GM inequality, we have
\begin{align*}
    E\left[(h(U,X)-Y)f'(X,W)\right] - \lambda \norm{f'}_2^2 =& E\left[E\left[h(U,X)-Y\mid X,W\right]f'(X,W)\right] - \lambda \norm{f'}_2^2 \\
    \leq& \norm{E\left[h(U,X)-Y\mid X,W\right]}_2\norm{f'}_2 - \lambda \norm{f'}_2^2\\
    =&\lambda \norm{f'}_2\left(\frac{\norm{E\left[h(U,X)-Y\mid X,W\right]}_2}{\lambda} - \norm{f'}_2\right)\\
    \leq & \frac{\norm{E\left[h(U,X)-Y\mid X,W\right]}_2^2}{4\lambda},
\end{align*}
in which the equality holds if and only if $f'(X,W) = E\left[h(U,X)-Y\mid X,W\right]/(2\lambda)$.
So we have
\begin{align*}
    \sup_{f'\in L^2(X,W)}  E\left[(h(U,X)-Y)f'(X,W)\right] - \lambda \norm{f'}_2^2 = \frac{\norm{E\left[h(U,X)-Y\mid X,W\right]}_2^2}{4\lambda} \geq 0,
\end{align*}
with in which the equality holds if and only if $E\left[h(U,X)\mid X,W\right] = E[Y\mid X,W]$, i.e., $h$ is a outcome bridge function. The second inequality can be showed similarly, as we observe that for a treatment bridge function $f_0$,
\begin{align*}
    E[f_0(X,W)h'(U,W)] =& E\left[E[f_0(X,W)\mid U,W]h'(U,W)\right]\\
    =&E\left[\frac{h'(U,W)}{\pi_W(U)}\right]\\
    =&E\left[\sum_{w=0,1}\frac{h'(U,w)I(W=w)}{\pi_w(U)}\right]\\
    =&E\left[\sum_{w=0,1}h'(U,w)\right].
\end{align*}
\subsection{Proof of Theorem \ref{asymptotich}}
To prove Theorems \ref{asymptotich} and \ref{asymptoticf}, we are using the following lemmas.
\begin{lemma}[Theorem 14.1, \cite{Wainwright2019high}]\label{lemma1}
    Given a star-shaped and $b$-uniformly bounded function class $\mathcal{G}$, let $\eta_n$ be any positive solution of the inequality $R_n(\eta; \mathcal{G}) \leq \eta^2 / b$. We call this solution the critical radius of $\mathcal{G}$. Then, for any $t \geq \eta_n$, we have
\begin{align*}
\abs{\norm{g}_n^2 - \norm{g}_2^2} \leq 0.5 \norm{g}_2^2 + 0.5t^2, \quad \forall g \in \mathcal{G},
\end{align*}
with probability greater than $1-c_1 \exp\left(-c_2\frac{nt^2}{b^2}\right)$.
\end{lemma}
Next, consider a function class $\mathcal{F} : X \to \mathbb{R}$ with loss $l : \mathbb{R} \times Z \to \mathbb{R}$.
\begin{lemma}[Lemma 14, \cite{Foster2023orthogonal}]\label{lemma2}
    Assume $\sup_{f \in \mathcal{F}} \| f \|_\infty \leq c$ and pick any $f^* \in \mathcal{F}$. Define $\eta_n$ be solution to
\begin{align*}
R_n(\eta; \operatorname{star}(\mathcal{F} - f^*)) \leq \eta^2 / c.
\end{align*}
Moreover, assume that the loss $l(\cdot, \cdot)$ is $L$-Lipschitz in the first argument. Then, for $\eta = \eta_n + \sqrt{c_0 \log(c_1 / \delta) / n}$ with some universal constants $c_0, c_1$, with $1 - \delta$,
\begin{align*}
\abs{ \left(\mathbb{E}_n[l(f(x), z)] - \mathbb{E}_n[l(f^*(x), z)]\right) - \left( \mathbb{E}[l(f(x), z)] - \mathbb{E}[l(f^*(x), z)] \right) } \lesssim L \eta_n (\| f - f^* \|_2 + \eta_n).
\end{align*}
\end{lemma}

Define
\begin{align*}
 &\Phi^\lambda(h, f') = E[h(U,W)f'(X,W)] - E[f'(X,W)Y] - \lambda \norm{f'}_2^2,\\
    &\Phi_n^\lambda(h, f')  = \hat{E}_a[h(U,W)f'(X,W)] - \hat{E}_m[f'(X,W)Y] - \lambda \norm{f'}_n^2,\\
    &\Phi(h,f') = \Phi^0(h,f'),\Phi_n(h,f') = \Phi_n^0(h,f'),
\end{align*}
where $\norm{f'}_n = \sqrt{\hat{E}_n[(f'(X,W))^2]}$ denotes the empirical $L^2$ norm with respect to the whole data.
Define $\eta_n = \eta_{h,n} + \sqrt{c_2\log(c_3/\delta)/n}$ and similarly define $\eta_{n_1},\eta_{n_2}$. Then, from Lemma \ref{lemma1}, we have
\begin{align*}
    \abs{\norm{f'}_n^2 - \norm{f'}_2^2} \leq  \frac12 \norm{f'}_2^2 + \frac{\eta_n^2}{2},
\end{align*} 
for all $f' \in F'$ with probability $1 - c_0\exp(-c_1n\eta_{n}^2/\norm{F'}_\infty^2)$. 
Equivalently, we have
\begin{align*}
    \frac12 \norm{f'}_2^2 - \frac{\eta_n^2}{2}\leq \norm{f'}_n^2  \leq \frac32 \norm{f'}_2^2 + \frac{\eta_n^2}{2}.
\end{align*}
Based on Lemma \ref{lemma2}, with probability $1-\delta/3$, we have
\begin{align*}
    \abs{\hat{E}_a[h(U,W)f'(X,W)] - E[h(U,W)f'(X,W)]} \leq C_1\left(\eta_{n_2}\norm{f'}_2 + \eta_{n_2}^2\right),
\end{align*}
where we let the loss function $l(f(x), z) = f(x)z$, $f = f'$, $z = h$,$f^*=0$ in the lemma.
Similarly, we have with probability $1-\delta/3$,
\begin{align*}
     \abs{\hat{E}_m[f'(X,W)Y] - E[f'(X,W)Y]} \leq C_1\left(\eta_{n_1}\norm{f'}_2 + \eta_{n_1}^2\right).
\end{align*}
To sum up, we have with probability $1-\frac23\delta$,
\begin{align*}
    \abs{\left(\hat{E}_a[h(U,W)f'(X,W)] - \hat{E}_m[f'(X,W)Y]\right) - \left(E[(h(U,W)-Y)f'(X,W)]\right)} \leq C_1\left((\eta_{n_1}+\eta_{n_2})\norm{f'}_2+(\eta_{n_1}^2+\eta_{n_2}^2)\right).
\end{align*}
Thus,
\begin{align*}
     \left(\hat{E}_a[h(U,W)f'(X,W)] - \hat{E}_m[f'(X,W)Y]\right)  \leq \left(E[(h(U,W)-Y)f'(X,W)]\right)+C_1\left((\eta_{n_1}+\eta_{n_2})\norm{f'}_2+(\eta_{n_1}^2+\eta_{n_2}^2)\right).
\end{align*}
\begin{align*}
    \sup_{f'\in F'}\Phi_n^\lambda(h_0,f') \leq& \sup_{f'\in F'}\left\{E[(h_0(U,W)-Y)f'(X,W)] + C_1\left((\eta_{n_1}+\eta_{n_2})\norm{f'}_2+(\eta_{n_1}^2+\eta_{n_2}^2)\right)-\frac\lambda2 \norm{f'}_2^2+\frac\lambda2 \eta_n^2 \right\}\\
    =&\sup_{f'\in F'}\left\{C_1\left((\eta_{n_1}+\eta_{n_2})\norm{f'}_2+(\eta_{n_1}^2+\eta_{n_2}^2)\right)-\frac\lambda2 \norm{f'}_2^2+\frac\lambda2 \eta_n^2 \right\}\\
    \leq&C_1\left(\eta_{n_1}^2 + \eta_{n_2}^2\right) + \frac{\lambda\eta_n^2}2 + \frac{C_1^2\left(\eta_{n_1} + \eta_{n_2}\right)^2}{2\lambda}.
\end{align*}
Moreover, 
\begin{align*}
    \sup_{f'\in F'} \Phi_n^\lambda(\hat{h},f') =& \sup_{f'\in F'} \left\{\Phi_n(\hat{h},f')-\Phi_n(h_0,f') + \Phi_n(h_0,f')-\lambda\norm{f'}_n^2\right\}\\
    \geq& \sup_{f'\in F'}\left\{\Phi_n(\hat{h},f')-\Phi_n(h_0,f') - 2\lambda\norm{f'}_n^2\right\}+\inf_{f'\in F'}\left\{ \Phi_n(h_0,f') + \lambda\norm{f'}_n^2\right\} \\
    =& \sup_{f'\in F'}\left\{\Phi_n(\hat{h},f')-\Phi_n(h_0,f') - 2\lambda\norm{f'}_n^2\right\}+\inf_{-f'\in F'}\left\{ \Phi_n(h_0,-f') + \lambda\norm{f'}_n^2\right\} \\
    =& \sup_{f'\in F'}\left\{\Phi_n(\hat{h},f')-\Phi_n(h_0,f') - 2\lambda\norm{f'}_n^2\right\}-\sup_{-f'\in F'}\left\{ \Phi_n(h_0,f') - \lambda\norm{f'}_n^2\right\} \\
    =& \sup_{f'\in F'}\left\{\Phi_n(\hat{h},f')-\Phi_n(h_0,f') - 2\lambda\norm{f'}_n^2\right\}-\sup_{f'\in F'} \Phi_n^\lambda(h_0,f').
\end{align*}
Then, we have
\begin{align*}
    \sup_{f'\in F'}\left\{\Phi_n(\hat{h},f')-\Phi_n(h_0,f') - 2\lambda\norm{f'}_n^2\right\} \leq& \sup_{f'\in F'} \Phi_n^\lambda(h_0,f')  + \sup_{f'\in F'}\Phi_n^\lambda(\hat{h},f')\\
    \leq& 2\sup_{f'\in  F'}\{\Phi_n^\lambda(h_0,f')\}\\
    \leq& 2C_1\left(\eta_{n_1}^2 + \eta_{n_2}^2\right) + \lambda\eta_n^2 + \frac{C_1^2\left(\eta_{n_1} + \eta_{n_2}\right)^2}{\lambda}.
\end{align*}
Let $f_{\hat{h}}(X,W) = E[\hat{h}(U,W)-h_0(U,W)\mid X,W] \in F'$. 
Suppose that $\norm{f_{\hat{h}}}_2\geq \eta_n$. 
Then, $rf_{\hat{h}} \in F'$ by the star-shape property of $F'$, where $r = \eta_n/\norm{f_{\hat{h}}}_2 \leq 1$. 
Hence, we have
\begin{align*}
    \sup_{f'\in F'}\left\{\Phi_n(\hat{h},f')-\Phi_n(h_0,f') - 2\lambda\norm{f'}_n^2\right\} \geq& \Phi_n(\hat{h},rf_{\hat{h}})-\Phi_n(h_0,rf_{\hat{h}}) - 2\lambda\norm{rf_{\hat{h}}}_n^2\\
    =&r \hat{E}_a[(\hat{h}(U,W)-h_0(U,W))f_{\hat{h}}(X,W)] - 2\lambda r^2 \norm{f_{\hat{h}}}_n^2.
\end{align*}
Again from Lemma \ref{lemma2}, where we let $f^* = 0$, $l(f(x),z)=f(x) z$ with $f = (\hat{h}-h_0)f_{\hat{h}}$ and $z = 1$, with probability $1-\delta/3$,
\begin{align*}
    \abs{\hat{E}_a[(\hat{h}(U,W) - h_0(U,W))f_{\hat{h}}(X,W)] - E[(\hat{h}(U,W) - h_0(U,W))f_{\hat{h}}(X,W)]}\leq &\eta_{n_2}\norm{(\hat{h}-h_0)f_{\hat{h}}}_2 + \eta_{n_2}^2\\
    \leq& C_1\eta_{n_2}(\norm{f_{\hat{h}}}_2 + \eta_{n_2}).
\end{align*}
And further,
\begin{align*}
    \hat{E}_a[(\hat{h}(U,W) - h_0(U,W))f_{\hat{h}}(X,W)] \geq& E[(\hat{h}(U,W) - h_0(U,W))f_{\hat{h}}(X,W)] - C_1\eta_{n_2}(\norm{f_{\hat{h}}}_2 + \eta_{n_2})\\
    &=\norm{f_{\hat{h}}}_2^2 - C_1\eta_{n_2}(\norm{f_{\hat{h}}}_2 + \eta_{n_2}).
\end{align*}
So, we have
\begin{align*}
    \sup_{f'\in F'}\left\{\Phi_n(\hat{h},f')-\Phi_n(h_0,f') - 2\lambda\norm{f'}_n^2\right\} \geq& r\left(\norm{f_{\hat{h}}}_2^2 - C_1\eta_{n_2}(\norm{f_{\hat{h}}}_2 + \eta_{n_2})\right) - 2\lambda r^2 \left(\frac32\norm{f_{\hat{h}}}_2^2+\frac12 \eta_n^2\right)\\
    \geq&\eta_n \norm{f_{\hat{h}}}_2 -C_1\eta_{n_2}\eta_n - C_1\eta_{n_2}^2 - 4\lambda \eta_n^2.
\end{align*}
To sum up, we have either $\norm{f_{\hat{h}}}_2 \leq \eta_n$, or that
\begin{align*}
    \norm{f_{\hat{h}}}_2 \leq C_1 \eta_{n_2} + \frac{C_1\eta_{n_2}^2}{\eta_n} + 5\lambda \eta_n + \frac{2C_1\left(\eta_{n_1}^2+\eta_{n_2}^2\right)}{\eta_n} + \frac{C_1^2(\eta_{n_1}+\eta_{n_2})^2}{\lambda \eta_n}.
\end{align*}
Therefore, with probability $1-c_0\exp(-c_1
n\eta_{n}^2/\norm{F'}_\infty^2)-\delta$, we have
\begin{align*}
    \norm{E[\hat{h}(U,W) - h_0(U,W)\mid X,W]}_2 = O\left(\left(1+\lambda+\frac1\lambda\right)\left(\eta_{h,n} + \sqrt{(1+\log(1/\delta))/n}\right)\right).
\end{align*}
\subsection{Proof of Theorem \ref{asymptotictauh}}
We have the following decomposition:
\begin{align}
    \hat\tau_w^h - \tau_w =& (\hat{E}_a - E)[h_0(U,w)] + (\hat{E}_a - E)[\hat{h}(U,w)-h_0(U,w)] + E[\hat{h}(U,w) - h_0(U,w)]  \notag\\
    \leq& \abs{(\hat{E}_a - E)[h_0(U,w)]} + \abs{(\hat{E}_a - E)[\hat{h}(U,w)-h_0(U,w)]} + \abs{E[\hat{h}(U,w) - h_0(U,w)]}\label{detauh}
\end{align}
The first term of (\ref{detauh}) converges to 0 in probability due to central limit theorem. 
The second term of (\ref{detauh}) can be bounded by
\begin{align*}
    \abs{(\hat{E}_a - E)[\hat{h}(U,w)-h_0(U,w)]} \leq \sup_{h \in H}\abs{(\hat{E}_a - E)[h(U,w)-h_0(U,w)]},
\end{align*}
which converges to 0 since $\{h(U,w)\mid h \in H\}$ is Glivenko-Cantelli class.
The third term of (\ref{detauh}) can be bounded by
\begin{align*}
    \abs{E\left[\hat{h}(U,w) - h_0(U,w)\right]} =& \abs{E\left[\left(\hat{h}(U,W) - h_0(U,W)\right)\frac{\I(W=w)}{\pi_w(U)}\right]}\\
    =& \abs{E\left[\left(\hat{h}(U,W) - h_0(U,W)\right)\I(W=w)E\left[f_0(X,W)\mid U,W\right]\right]}\\
    =& \abs{E\left[\left(\hat{h}(U,W)-h_0(U,W)\right)\I(W=w)f_0(X,W)\right]}\\
    =&\abs{E\left[E\left[\hat{h}(U,W)-h_0(U,W)\mid X,W\right]\I(W=w)f_0(X,W)\right]}\\
    \leq& \norm{E[\hat{h}(U,W)-h_0(U,W)\mid X,W]}_2 \norm{\I(W=w)f_0(X,W)}_2,
\end{align*}
which converges based on the results of Theorem \ref{asymptotich}. 
Therefore, the estimator $\hat\tau_w^h$ is consistent.

\subsection{Proof of Theorem \ref{asymptoticf}}
The proof of Theorem \ref{asymptoticf} is similar to Theorem \ref{asymptotich}. Define 
\begin{align*}
    &\Phi^{\lambda}(f,h') = E[f(X,W)h'(U,W)-h'(U,0)-h'(U,1)] - \lambda \norm{h'}^2,\\
    &\Phi_n^{\lambda} (f,h') = \hat{E}_a \left[f(X,W)h'(U,W) -h'(U,0)-h'(U,1)\right] - \lambda \norm{h'}_a^2,\\
    &\Phi(f,h') = \Phi^0(f,h'),\Phi_n(f,h') = \Phi_n^0(f,h'),  
\end{align*}
where $\norm{h'}_a = \sqrt{\hat{E}_a[(h'(U,W))^2]}$ denotes the empirical $L^2$ norm with respect to the auxiliary data.
Define $\eta_{n_2} = \eta_{f,n_2} + \sqrt{c_2\log(c_3/\delta)/n_2}$. 
Then, from Lemma \ref{lemma1}, we have
\begin{align*}
    \abs{\norm{h'}_a^2 - \norm{h'}_2^2} \leq \frac12 \norm{h'}_2^2 + \frac{\eta_{n_2}^2}{2},
\end{align*}
or,
\begin{align*}
    \frac12 \norm{h'}_2^2 - \frac{\eta_{n_2}^2}{2} \leq \norm{h'}_a^2 \leq \frac32 \norm{h'}_2^2 + \frac{\eta_{n_2}^2}{2},
\end{align*}
for all $h' \in H'$ with probability $1-c_0 \exp(-c_1n_2 \eta_{n_2}^2/\norm{H'}_\infty^2)$.
From Lemma \ref{lemma2}, where we let $f^*=0$, $l(f(x), z) = f(x)z$, and $f$ and $z$ in the lemma are $h'$ and $f$ here, with probability $1-\delta/3$, we have
\begin{align*}
\abs{\hat{E}_a[f(X,W)h'(U,W)] - E[f(X,W)h'(U,W)]} \leq C_1\left(\eta_{n_2}\norm{h'}_2 +\eta_{n_2}^2\right)/2.
\end{align*}
Similarly, with probability $1-\delta/3$, we have
\begin{align*}
    \abs{\hat{E}_a[h'(U,1) + h'(U,0)] - E[h'(U,1) + h'(U,0)]} \leq& \eta_{n_2}\norm{h'(U,1)+h'(U,0)}_2 + \eta_{n_2}^2\\
    \leq& C_1(\eta_{n_2}\norm{h'}_2 + \eta_{n_2}^2)/2.
\end{align*}
To sum up, we have
\begin{align*}
    \abs{\hat{E}_a[f(X,W)h'(U,W)-h'(U,1)-h'(U,0)]-E[f(X,W)h'(U,W)-h'(U,1)-h'(U,0)]} \leq C_1\left(\eta_{n_2}\norm{h'}_2 +\eta_{n_2}^2\right),
\end{align*}
and further
\begin{align*}
    \hat{E}_a[f(X,W)h'(U,W)-h'(U,1)-h'(U,0)] \leq E[f(X,W)h'(U,W)-h'(U,1)-h'(U,0)] + C_1\left(\eta_{n_2}\norm{h'}_2 +\eta_{n_2}^2\right).
\end{align*}
Thus, 
\begin{align*}
    \sup_{h'\in H'} \Phi_n^\lambda(f_0, h') \leq& \sup_{h'\in H'} \left\{ E[f_0(X,W)h'(U,W)-h'(U,1)-h'(U,0)] +C_1\left(\eta_{n_2}\norm{h'}_2 + \eta_{n_2}^2\right) - \frac{\lambda}{2}\norm{h'}_2^2 + \frac{\lambda}{2}\eta_{n_2}^2\right\}\\
    =& \sup_{h'\in H'}\left\{C_1\left(\eta_{n_2}\norm{h'}_2 + \eta_{n_2}^2\right) - \frac{\lambda}{2}\norm{h'}_2^2 + \frac{\lambda}{2}\eta_{n_2}^2\right\}\\
    \leq& \left(C_1 + \frac{\lambda}{2} + \frac{C_1^2}{2\lambda}\right) \eta_{n_2}^2.
\end{align*}
Moreover,
\begin{align*}
    \sup_{h'\in H'} \Phi_n^\lambda(\hat{f},h') =& \sup_{h'\in H'} \left\{\Phi_n(\hat{f},h')-\Phi_n(f_0,h')+\Phi_n(f_0,h')-\lambda\norm{h'}_a^2\right\}\\
    \geq& \sup_{h' \in H'} \left\{\Phi_n(\hat{f},h') - \Phi_n(f_0,h') - 2\lambda\norm{h'}_a^2\right\} + \inf_{h' \in H'}\left\{\Phi_n(f_0,h') + \lambda \norm{h'}_a^2\right\}\\
    =& \sup_{h' \in H'} \left\{\Phi_n(\hat{f},h') - \Phi_n(f_0,h') - 2\lambda\norm{h'}_a^2\right\} + \inf_{-h' \in H'}\left\{\Phi_n(f_0,-h') + \lambda \norm{h'}_a^2\right\}\\
    =&\sup_{h' \in H'} \left\{\Phi_n(\hat{f},h') - \Phi_n(f_0,h') - 2\lambda\norm{h'}_a^2\right\} - \sup_{-h' \in H'}\left\{\Phi_n(f_0,h') - \lambda \norm{h'}_a^2\right\}\\
    =&\sup_{h' \in H'} \left\{\Phi_n(\hat{f},h') - \Phi_n(f_0,h') - 2\lambda\norm{h'}_a^2\right\} - \sup_{h' \in H'}\Phi_n^\lambda(f_0,h').
\end{align*}
Then, we have
\begin{align*}
    \sup_{h' \in H'} \left\{\Phi_n(\hat{f},h') - \Phi_n(f_0,h') - 2\lambda\norm{h'}_a^2\right\} \leq& \sup_{h'\in H'} \Phi_n^\lambda(\hat{f},h') + \sup_{h' \in H'}\Phi_n^\lambda(f_0,h')\\
    \leq& 2 \sup_{h' \in H'}\Phi_n^\lambda(f_0,h')\\
    \leq& \left(2C_1 + \lambda + \frac{C_1^2}{\lambda}\right) \eta_{n_2}^2.
\end{align*}
Let $h_{\hat{f}}(U,W) = E[\hat{f}(X,W)-f_0(X,W)\mid U,W] \in H'$. Suppose that $\norm{h_{\hat{f}}}_2 \geq \eta_{n_2}$. 
Then, $rh_{\hat{f}} \in H'$ by the star-shape property of $H'$, where $r = \eta_{n_2}/\norm{h_{\hat{f}}}_2\leq 1$. 
Hence, we have
\begin{align*}
    \sup_{h' \in H'} \left\{\Phi_n(\hat{f},h') - \Phi_n(f_0,h') - 2\lambda\norm{h'}_a^2\right\} \geq& \Phi_n(\hat{f},rh_{\hat{f}}) - \Phi_n(f_0,rh_{\hat{f}}) - 2\lambda \norm{rh_{\hat{f}}}_a^2\\
    =& r\hat{E}_a\left[ \left(\hat{f}(X,W)-f_0(X,W)\right)h_{\hat{f}}(U,W)\right]-2\lambda r^2 \norm{h_{\hat{f}}}_a^2.
\end{align*}
Again from Lemma \ref{lemma2} where we let $f^*=0$, $l(f(x),z)=f(x)z$, $f = (\hat{f}-f_0)h_{\hat{f}}$, $z = 1$, with probability $1-\delta/3$,
\begin{align*}
    \abs{\hat{E}_a\left[ \left(\hat{f}(X,W)-f_0(X,W)\right)h_{\hat{f}}(U,W)\right]-E\left[ \left(\hat{f}(X,W)-f_0(X,W)\right)h_{\hat{f}}(U,W)\right]} \leq & \eta_{n_2}\norm{(\hat{f}-f_0)h_{\hat{f}}}_2 + \eta_{n_2}^2\\
    \leq& C_1 \eta_{n_2}\left(\norm{h_{\hat{f}}}_2 + \eta_{n_2}\right),
\end{align*}
and further,
\begin{align*}
    \hat{E}_a\left[ \left(\hat{f}(X,W)-f_0(X,W)\right)h_{\hat{f}}(U,W)\right] \geq& E\left[ \left(\hat{f}(X,W)-f_0(X,W)\right)h_{\hat{f}}(U,W)\right] - C_1 \eta_{n_2}\left(\norm{h_{\hat{f}}}_2 + \eta_{n_2}\right)\\
    =& \norm{h_{\hat{f}}}_2^2 - C_1 \eta_{n_2}\left(\norm{h_{\hat{f}}}_2+\eta_{n_2}\right).
\end{align*}
So, we have
\begin{align*}
    \sup_{h' \in H'} \left\{\Phi_n(\hat{f},h') - \Phi_n(f_0,h') - 2\lambda\norm{h'}_a^2\right\} \geq& r\left(\norm{h_{\hat{f}}}_2^2 - C_1 \eta_{n_2}\left(\norm{h_{\hat{f}}}_2+\eta_{n_2}\right)\right)-2\lambda r^2\left(\frac32 \norm{h_{\hat{f}}}_2^2 + \frac12 \eta_{n_2}^2\right)\\
    \geq& \eta_{n_2}\norm{h_{\hat{f}}}_2 - 2C_1\eta_{n_2}^2 - 4\lambda \eta_{n_2}^2.
\end{align*}
To sum up, we have either $\norm{h_{\hat{f}}}_2 \leq \eta_{n_2}$, or that
\begin{align*}
    \norm{h_{\hat{f}}}_2 \leq (4C_1+5\lambda+\frac{C_1^2}{\lambda}) \eta_{n_2}.
\end{align*}
Therefore, with probability $1-c_0 \exp(-c_1n_2 \eta_{n_2}^2/\norm{H'}_\infty^2)-\delta$, we have
\begin{align*}
    \norm{E[\hat{f}(X,W)-f_0(X,W)\mid U,W]}_2 = O\left(\left(1+\lambda+\frac1\lambda\right)\left(\eta_{h,n_2}+\sqrt{(1+\log(1/\delta))/n_2}\right)\right) .
\end{align*}

\subsection{Proof of Theorem \ref{asymptotictauf}}
We have the following decomposition:
\begin{align}
    \hat\tau_w^f - \tau_w =& (\hat{E}_m-E)\left[f_0(X,W)\I(W=w)Y\right] + (\hat{E}_m-E)\left[(\hat{f}(X,W)-f_0(X,W))\I(W=w)Y\right]\notag \\&+ E\left[(\hat{f}(X,W)-f_0(X,W))\I(W=w)Y\right]\notag\\
    \leq&\abs{(\hat{E}_m-E)\left[f_0(X,W)\I(W=w)Y\right]} + \abs{(\hat{E}_m-E)\left[(\hat{f}(X,W)-f_0(X,W))\I(W=w)Y\right]}\notag\\
    +&\abs{E\left[(\hat{f}(X,W)-f_0(X,W))\I(W=w)Y\right]}.\label{detauf}
\end{align}
The first term of (\ref{detauf}) converges to 0 in probability due to central limit theorem. The second term of (\ref{detauf}) can be bounded by
\begin{align*}
\abs{(\hat{E}_m-E)\left[(\hat{f}(X,W)-f_0(X,W))\I(W=w)Y\right]} \leq \sup_{f\in F}\abs{(\hat{E}_m-E)\left[(f(X,W)-f_0(X,W))\I(W=w)Y\right]}, 
\end{align*}
which converges to 0 since $F$ is a Glivenko-Cantelli class. The third term of $\ref{detauf}$ can be bounded by
\begin{align*}
    \abs{E\left[(\hat{f}(X,W)-f_0(X,W))\I(W=w)Y\right]}
    =&\abs{E\left[(\hat{f}(X,W)-f_0(X,W))\I(W=w)E\left[Y\mid X,W\right]\right]}\\
    =&\abs{E\left[(\hat{f}(X,W)-f_0(X,W))\I(W=w)h_0(U,X)\right]}\\
    =&\abs{E\left[E\left[\hat{f}(X,W)-f_0(X,W)\mid U,W\right]\I(W=w)h_0(U,W)\right]}\\
    \leq& \norm{E\left[\hat{f}(X,W)-f_0(X,W)\mid U,W\right]}_2 \norm{\I(W=w)h_0(U,W)}_2,
\end{align*}
which converges based on the results of Theorem \ref{asymptoticf}. Therefore, the estimator $\hat\tau_w^f$ is consistent.
\subsection{Proof of Theorem \ref{asymptoticdb}}
We have the following decomposition:
\begin{align*}
    \hat\tau_w^\text{dr} - \tau_w =& (\hat{E}_m - E)\left[f_0(X,W)\I(W=w)Y\right] + (\hat{E}_a - E)\left[\left(1-f_0(X,W)\I(W=w)\right)h_0(U,w)\right]\\
    &+E\left[\left(\hat{f}(X,W)-f_0(X,W)\right)\I(W=w)Y\right] + E\left[\left(1-\hat{f}(X,W)\I(W=w)\right)\hat{h}(U,w)\right]\\
    &+(\hat{E}_m - E)\left[\left(\hat{f}(X,W)-f_0(X,W)\right)\I(W=w)Y\right]\\
    &+(\hat{E}_a - E)\left[\left(1-\hat{f}(X,W)\I(W=w)\right)\hat{h}(U,w)-\left(1-f_0(X,W)\I(W=w)\right)h_0(U,w) \right].
\end{align*}
Since $\hat{h},\hat{f}$ belong to Donsker classes, we have
\begin{align*}
    &(\hat{E}_m - E)\left[\left(\hat{f}(X,W)-f_0(X,W)\right)\I(W=w)Y\right]\\
    &+(\hat{E}_a - E)\left[\left(1-\hat{f}(X,W)\I(W=w)\right)\hat{h}(U,w)-\left(1-f_0(X,W)\I(W=w)\right)h_0(U,w) \right] = o_P(1).
\end{align*}
We observe that,
\begin{align*}
    &E\left[\left(\hat{f}(X,W)-f_0(X,W)\right)\I(W=w)Y\right] = E\left[\left(\hat{f}(X,W)-f_0(X,W)\right)h_0(U,W)\I(W=w)\right],\\
    &E\left[\hat{h}(U,w)-h_0(U,w)\right]=E\left[\left(\hat{h}(U,W)-h_0(U,W)\right)\I(W=w)f_0(X,W)\right],\\
    &E\left[\left(1-f_0(X,W)\I(W=w)\right)h_0(U,w)\right] = 0.
\end{align*}
Therefore, we have
\begin{align*}
    &\abs{E\left[\left(\hat{f}(X,W)-f_0(X,W)\right)\I(W=w)Y\right] + E\left[\left(1-\hat{f}(X,W)\I(W=w)\right)\hat{h}(U,w)\right]}\\ =& \abs{E\left[\left(f_0(X,W)-\hat{f}(X,W)\right)\left(\hat{h}(U,W)-h_0(U,W)\right)\I(W=w)\right]}\\
    =& \abs{E\left[E\left[f_0(X,W)-\hat{f}(X,W)\mid U,W\right]\left(\hat{h}(U,W)-h_0(U,W)\right)\I(W=w)\right]}\\
    =&\abs{E\left[\left(f_0(X,W)-\hat{f}(X,W)\right)E\left[\hat{h}(U,W)-h_0(U,W)\mid X,W\right]\I(W=w)\right]}\\
    \leq & \min\left\{\norm{E\left[f_0(X,W)-\hat{f}(X,W)\mid U,W\right]}_2\norm{\hat{h}-h_0}_2 ,\norm{E\left[\hat{h}(U,W)-h_0(U,W)\mid X,W\right]}_2\norm{\hat{f}-f_0}_2\right\}.
\end{align*}
Thus, we obtain that $\hat\tau_w^\text{dr}$ is consistent if either $\norm{E\left[f_0(X,W)-\hat{f}(X,W)\mid U,W\right]}_2$ or $\norm{E\left[\hat{h}(U,W)-h_0(U,W)\mid X,W\right]}_2$ converges to 0 in probability.

If both  $\norm{E\left[f_0(X,W)-\hat{f}(X,W)\mid U,W\right]}_2$ and $\norm{E\left[\hat{h}(U,W)-h_0(U,W)\mid X,W\right]}_2$ converge, we have
\begin{align*}
    &(\hat{E}_m - E)\left[\left(\hat{f}(X,W)-f_0(X,W)\right)\I(W=w)Y\right]\\
    &+(\hat{E}_a - E)\left[\left(1-\hat{f}(X,W)\I(W=w)\right)\hat{h}(U,w)-\left(1-f_0(X,W)\I(W=w)\right)h_0(U,w) \right] = O_P(n^{-1/2}).
\end{align*}
due to Donsker class property.
Besides, 
\begin{align*}
    &\abs{E\left[\left(\hat{f}(X,W)-f_0(X,W)\right)\I(W=w)Y\right] + E\left[\left(1-\hat{f}(X,W)\I(W=w)\right)\hat{h}(U,w)\right]} \\
    \leq& \norm{E\left[\hat{h}(U,W)-h_0(U,W)\mid X,W\right]}_2\norm{\hat{f}-f_0}_2\\
    \leq& c^F\norm{E\left[\hat{h}(U,W)-h_0(U,W)\mid X,W\right]}_2\norm{E\left[f_0(X,W)-\hat{f}(X,W)\mid U,W\right]}_2;\\\\
    &\abs{E\left[\left(\hat{f}(X,W)-f_0(X,W)\right)\I(W=w)Y\right] + E\left[\left(1-\hat{f}(X,W)\I(W=w)\right)\hat{h}(U,w)\right]} \\
    \leq& \norm{E\left[f_0(X,W)-\hat{f}(X,W)\mid U,W\right]}_2\norm{\hat{h}-h_0}_2\\
    \leq& c^H \norm{E\left[\hat{h}(U,W)-h_0(U,W)\mid X,W\right]}_2\norm{E\left[f_0(X,W)-\hat{f}(X,W)\mid U,W\right]}_2,
\end{align*}
The last inequalities hold since we can choose a $f_0 \in F$ and $h_0 \in H$ such that
\begin{align*}
    \frac{\norm{\hat{f}-f_0}_2}{\norm{E\left[f_0(X,W)-\hat{f}(X,W)\mid U,W\right]}_2} < c^F,
    \frac{\norm{\hat{h}-h_0}_2}{\norm{E\left[\hat{h}(U,W)-h_0(U,W)\mid X,W\right]}_2} < c^H,
\end{align*}
by the definition of the ill-posedness measures.
So we have
\begin{align*}
        \hat\tau_w^\text{dr} - \tau_w =O_P\left(\min\{c^H,c^F\} ||E[\hat{h}(U,W)-Y|X,W]||_2\cdot||E[\hat{f}(X,W)-f_0(X,W)|U,W]||_2+n^{-1/2}\right). 
    \end{align*}

Further, if there is a unique treatment bridge function $f_0 \in F$ and a unique outcome bridge function $h_0 \in H$, then we have
\begin{align*}
    \norm{\left(\hat{f}(X,W) - f_0(X,W)\right)I(W=w)Y}_2 = O_P\left(c^F \norm{E\left[\hat{f}(X,W)-f_0(X,W)\mid U,W\right]}_2\right).
\end{align*} 
Therefore, if $c^F < \infty$ and $\norm{E\left[\hat{f}(X,W)-f_0(X,W)\mid U,W\right]}_2 = o_p(1)$, then we have for any small $\epsilon ,\delta>0$, with probability $1-\delta$, $\norm{\hat{f}(X,W) - f_0(X,W)}_2 < \epsilon$ for a large enough $n$.
Due to the Donsker class property, $\sqrt{n_1}(\hat{E}_m-E)\left[(f-f_0)I(W=w)Y\right]$ converges to a Gaussian process, denoted by $G\left[(f-f_0)I(W=w)Y\right]$ in distribution.
Then,
\begin{align*}
    &\abs{\sqrt{n_1}(\hat{E}_m - E)\left[\left(\hat{f}(X,W)-f_0(X,W)\right)I(W=w)Y\right]}\\ \leq& \sup\limits_{\norm{f(X,W) - f_0(X,W)}_2 < \epsilon}\abs{ \sqrt{n_1}(\hat{E}_m - E)\left[\left(f(X,W)-f_0(X,W)\right)I(W=w)Y\right]}\\
    \to&\sup\limits_{\norm{f(X,W) - f_0(X,W)}_2 < \epsilon} \abs{G\left[\left(f(X,W)-f_0(X,W)\right)I(W=w)Y\right]},
\end{align*}
which converges to zero based on Corollary 1.2 in \cite{Kraetschmer2023Kolmogorov}.
Therefore, we have $(\hat{E}_m - E)\left[\left(\hat{f}(X,W)-f_0(X,W)\right)\I(W=w)Y\right] = o_P(n^{-1/2})$.
Similarly, if further $c^H < \infty$ and $\norm{E\left[\hat{h}(U,W)-h_0(U,W)\mid X,W\right]}=o_P(1)$, we can show that
\begin{align*}
    \norm{\left(1-\hat{f}(X,W)\I(W=w)\right)\hat{h}(U,w)-\left(1-f_0(X,W)\I(W=w)\right)h_0(U,w)}_2 = o_P(1),
\end{align*}
and hence
\begin{align*}
    (\hat{E}_a - E)\left[\left(1-\hat{f}(X,W)\I(W=w)\right)\hat{h}(U,w)-\left(1-f_0(X,W)\I(W=w)\right)h_0(U,w) \right] = o_P(n^{-1/2}).
\end{align*}
In summary, we have shown the results of the theorem.
\subsection{Existence of the treatment bridge function}
We denote $\mathcal{L}: L^2(X,W)\mapsto L^2(U,W)$ as the operator of conditional expectation, $\mathcal{L}f = E[f(X,W)\mid U,W]$, and its adjoint operator $\mathcal{L^*}: L^2(X,W)\mapsto L^2(U,W)$, i.e, $\mathcal{L^*}h = E[h(U,W)\mid X,W]$. The treatment bridge function $f_0$ satisfies $\mathcal{L}f_0 = 1/\pi_W(U)$.
Suppose $(\lambda_n, \varphi_n, \psi_n)$ is the singular value decomposition for $\mathcal{L}$, we have the following proposition that ensures the existence of the treatment bridge function: 
\begin{proposition}
\label{conditionfortbridgefct}
The treatment bridge function exists if Assumption \ref{strictpositivity} holds and further
\begin{itemize}
    \item[(1)] $\int_u\int_x p_{X\mid U,W}(x\mid u,w)p_{U\mid X,W}(u\mid x,w) dxdu < +\infty  $;
    \item[(2)] $\mathcal{N}(\mathcal{L^\ast}) = 0$, which is called completeness condition;
    \item[(3)] $\sum_{i=1}^{+\infty} \lambda_n^{-2} |E(\psi_n/\pi_W(U))| < +\infty$.
\end{itemize}
\end{proposition}
\textit{Proof of Proposition \ref{conditionfortbridgefct}.} We apply Lemma \ref{picardthm} to prove Proposition \ref{conditionfortbridgefct}.
\begin{lemma}[Picard's Theorem]\label{picardthm} Letting  $K : H_1 \longmapsto H_2$ be a compact operator with singular system $(\lambda_n, \varphi_n, \psi_n)_{n=1}^{+\infty}$, where $\varphi_n \in H_1, \psi_n \in H_2$. Given $\phi \in H_2$, the equation of the first kind $Kh = \phi$ is solvable if and only if
\begin{enumerate}
    \item $\phi \in \mathcal{N}(K^*)^{\perp}$; and
    \item $\sum_{n=1}^{+\infty} \lambda_n^{-2} \left| \langle \phi, \psi_n \rangle \right|^2 < +\infty$; 
\end{enumerate}
where \( \mathcal{N}(K^*) = \{ h : K^* h = 0 \} \) is the null space of \( K^*, \) and \( \perp \) denotes the orthogonal complement to a subset.
\end{lemma}
First, we note that $\mathcal{L}$ is a compact operator by assuming $\int_u\int_x p_{X\mid U,W}(x\mid u,w)p_{U\mid X,W}(u\mid x,w) dxdu < +\infty$ \citep{Carrasco2007Linear}.
Thus, there exists a singular value decomposition $(\lambda_n, \varphi_n, \psi_n)_{n=1}^{+\infty}$ of $\mathcal{L}$ according to \cite{Kress1989Linear} and \cite{Carrasco2007Linear}.
Second, under the completeness condition (2), we have $1/\pi_W(U) \in  \mathcal{N}(\mathcal{L}^*)^{\perp} = L^2(U,W)$.
Therefore, the conditions for Lemma \ref{picardthm} holds, which implies the existence of the treatment bridge function.

Now, we give an example for the distribution of $(U,X,W)$, in which a treatment bridge function exists. 
Suppose that $U\in \{0,1\}$ obeys the Bernoulli distribution, $X = U + \varepsilon$ where $\epsilon \ind U$ and $\epsilon$ follows a standard normal distribution and $W = I(X \geq 0)$. 
In this case, we have
\begin{align*}
    &p_{W\mid U}(W\mid U) = W\Phi(U) + (1-W)(1-\Phi(U)), \\
    &p_{X\mid U,W}(X\mid U,W) = \frac{\phi(X-U)(WI(X \geq 0)+(1-W)I(X<0))}{W\Phi(U) + (1-W)(1-\Phi(U))},
\end{align*}
where $\phi,\Phi$ are density function and distribution function of a normal standard distribution, respectively. Then, a sufficient and necessary condition for the bridge function $f$ is, for $u \in \{0,1\}$ 
\begin{align*}
    \int_0^\infty f(x,1)\phi(x-u) dx = 1;\\
    \int_{-\infty}^0 f(x,0)\phi(x-u) dx = 1.
\end{align*}
Let $\tilde{f}(x,w)$ be a non-trivial solution to
\begin{align*}
\begin{cases}
    \int_0^\infty \tilde{f}(x,1) (\phi(x) - \phi(x-1))dx = 0,\\
    \int_{-\infty}^0 \tilde{f}(x,0)(\phi(x)-\phi(x-1)) dx = 0.
    \end{cases}
\end{align*}
Then, a bridge function can be derived as $f(x,1) = \tilde{f}(x,1)/\int_0^\infty \tilde{f}(x,1)\phi(x)dx, f(x,0) = \tilde{f}(x,0)/\int_{-\infty}^0 \tilde{f}\phi(x)dx$. 
\subsection{Details for Simulations}
\paragraph{Models.}
In our experiments, we use a two-layer neural network with the ReLu activation function to model $h(u,w)$ and $f(x,w)$. The input size is 2, the hidden size is 10, and the output size is 1. The basis functions of $H$ and $F$ are cosine functions, i.e., 
$$
\phi(x,w) = [\cos(i\cdot x) + w \cdot \cos(i\cdot x)]_{i=1}^{d_1}, 
$$
and
$$
\psi(u,w) = [\cos(i\cdot u) + w \cdot \cos(i\cdot u)]_{i=1}^{d_2}.
$$

\paragraph{Hyperparameters.}
We set $\gamma_1 = \gamma_2 = 0.03$, $\lambda_1 = \lambda_2 = 1.0$, $d_1 = d_2 = 10$.

\paragraph{Training.}
We train $h$, and $f$ as follows. In setting 1, we run 100 epochs, leverage the gradient descent method with Adam optimizer, and set the learning rate to 0.05. In setting 2, we run 100 epochs, leverage gradient descent method with Adam optimizer, and set the learning rate to 0.1. 
For bootstrapping, we repeat this 1000 times. 

\paragraph{Model Misspecification.} To verify the doubly robustness of $\hat\tau_w^\text{db}$, we provide  ``bad'' estimates for the treatment and outcome bridge functions, respectively, by specifying constant function space for constructing the bridge functions.

\end{document}